%% file: main.tex
\documentclass[fleqn,usenatbib]{mnras}

\usepackage{mathptmx}

\usepackage[T1]{fontenc}

\usepackage{graphicx}
\usepackage{amsmath}
\usepackage{amssymb}
\usepackage{bm}
\usepackage{mathtools}
\usepackage{siunitx}
\usepackage{subcaption}
\usepackage[acronym]{glossaries}

\usepackage[thinc]{esdiff}

\newcommand{\figref}[1]{\mbox{Figure}~\ref{#1}}
\newcommand{\secref}[1]{\mbox{Section}~\ref{#1}}
\newcommand{\tabref}[1]{\mbox{Table}~\ref{#1}}
\renewcommand{\eqref}[1]{Eq.~\ref{#1}}
\newcommand{\msol}{M_\odot}
\newcommand{\vel}{{\rm v}}
\newcommand{\velvec}{{\rm \mathbf{v}}}
\renewcommand{\vec}[1]{\mathbf{#1}}

\newacronym{igw}{IGW}{internal gravity wave}
\newacronym{ms}{MS}{main-sequence}
\newacronym{zams}{ZAMS}{zero-age main-sequence}
\newacronym{iles}{ILES}{implicit large eddy simulation}

\title[IGW Mixing in Stars: Simulations And Theory]{Mixing by Internal Gravity Waves in Stars: Assessing Numerical Simulations Against Theory}

\author[J. Morton et al.]{
J. Morton,$^{1}$\thanks{E-mail: j.j.morton@exeter.ac.uk}
T. Guillet,$^{1}$
I. Baraffe,$^{1, 2}$
A. Morison,$^{1}$
A. Le Saux,$^{3}$
D.G. Vlaykov,$^{4}$
T. Goffrey,$^{5, 6}$
and J. Pratt$^{7}$
\\
$^{1}$Physics and Astronomy, University of Exeter, Exeter, EX4 4QL, UK\\
$^{2}$Ecole Normale Sup\'erieure, Lyon, CRAL (UMR CNRS 5574), Universit\'e de Lyon, France\\
$^{3}$Université Paris-Saclay, Université Paris Cité, CEA, CNRS, AIM, 91191, Gif-sur-Yvette, France\\
$^{4}$Mathematics and Statistics, University of Exeter, Exeter EX4 4QF, UK\\
$^{5}$Centre for Fusion, Space and Astrophysics, Department of Physics, University of Warwick, Coventry CV4 7AL, UK\\
$^{6}$Healthcare Technology Institute, School of Chemical Engineering, The University of Birmingham, Birmingham, B15 2TT, UK\\
$^{7}$Lawrence Livermore National Laboratory, 7000 East Ave, Livermore, CA 94550, USA
}

\date{Accepted XXX. Received YYY; in original form ZZZ}
\pubyear{2025}

\begin{document}\label{firstpage}
\pagerange{\pageref{firstpage}--\pageref{lastpage}}
\maketitle

\begin{abstract}
\input{src/abstract}
\end{abstract}
\glsresetall


\input{src/01-intro}
\input{src/02-models}
\input{src/03-waves}
\input{src/04-mixing}
\input{src/05-tracers}

\input{src/06-discussion}

\section*{Acknowledgements}

The authors are thankful to H. Spruit for valuable discussions.
I. Baraffe thanks the Max Planck Institute for Astrophysics (MPA), Garching for hospitality during completion of this work.
The authors thank the anonymous referee for their comments which improved the manuscript.
This work is supported by the ERC grant No. 787361-COBOM and the STFC Consolidated Grant ST/V000721/1.
The authors would like to acknowledge the use of the University of Exeter High Performance Computing (HPC) facility ISCA and of the DiRAC Data Intensive service at Leicester, operated by the University of Leicester IT Services, which forms part of the STFC DiRAC HPC Facility.
The equipment was funded by BEIS capital funding via STFC capital grants ST/K000373/1 and ST/R002363/1 and STFC DiRAC Operations grant ST/R001014/1.
DiRAC is part of the National e-Infrastructure.
Part of this work was performed under the auspices of the U.S. Department of Energy by Lawrence Livermore National Laboratory under Contract DE-AC52-07NA27344 (LLNL-JRNL-867493).

\section*{Data Availability}

Data backing this study will be shared on reasonable request to the corresponding author.


\bibliographystyle{mnras}
\bibliography{ref}

\appendix
\include{src/appendix}

\bsp
\label{lastpage}
\end{document}

%% file: src/abstract.tex

Here we present a study of radial chemical mixing in non-rotating massive main-sequence stars driven by \glspl*{igw}, based on multi-dimensional hydrodynamical simulations with the fully compressible code MUSIC.
We examine two proposed mechanisms of material mixing in stars by \glspl*{igw} that are commonly quoted, relating to thermal diffusion and sub-wavelength shearing.
Thermal diffusion provides a non-restorative effect to the waves, leaving material displaced from its previous equilibrium, while shearing arising within the waves drives weak localised flows, mixing the fluid there.
Using \gls*{igw} spectra from the simulations, we evaluate theoretical predictions of mixing rates due to these mechanisms.
We show, for $20M_\odot$ main-sequence stars, that neither of these mechanisms are likely to create mixing sufficient to correct inaccuracies in current stellar evolution models.
Furthermore, we compare these predictions to results obtained from Lagrangian tracer particles, following a method recently used for global simulations of stellar interiors to measure mixing by \glspl*{igw} in their radiative zones.
We demonstrate that tracer particle methods face significant numerical challenges in measuring the small diffusion coefficients predicted by the aforementioned theories, for which they are prone to yielding artificially enhanced coefficients.
Diffusion coefficients based on such methods are currently used with stellar evolution codes for asteroseismic studies, but should be viewed with caution.
Finally, in a case where tracer particles do not suffer from numerical artefacts, we suggest that a diffusion model is not suitable for timescales typically considered by two-dimensional numerical simulations.

%% file: src/01-intro.tex

\section{Introduction}\label{sec:intro}

Mixing through the stably stratified regions of stars is critical to their evolution.
Understanding the processes involved is needed to correctly calibrate stellar evolutionary tracks and interpret abnormal surface abundances.
The existence of chemical mixing throughout radiative zones of stars is clear from observations of chemical under- and over-abundances at the surface.
For example, lithium depletion in metal-poor giants \citep{pilachowski93,palacios2006} and nitrogen over-abundances in massive \gls*{ms} stars \citep{brott2011} imply mixing between the core and observable surface.
Recently, asteroseismic studies \citep{pedersen21,burssens23} suggest that convective boundary mixing is also required to explain enlarged cores of intermediate- and high-mass \gls*{ms} stars.
However, the mechanisms acting to mix both material and angular momentum radially throughout stably stratified radiative regions \citep{pinsonneault97} are still unclear, and discrepancies often arise between models and inferences from observations \citep[e.g.][]{castro14,brott2011}.

Mechanisms with potential to give rise to mixing include convective overshooting and penetration, shearing, rotation, meridional flows, and \glspl*{igw}.
None of these processes have been shown to dominate in all stellar radiative regions, so the complete picture likely requires considering the effects of each \citep{johnston21,pedersen21}.
Convective overshooting describes convective plumes which reach ballistically through the convective-radiative boundary, acting to transport both material and angular momentum into the radiative zone.
\citet{castro14} suggested that models underestimate the mixing due to overshooting, a discrepancy that increases with stellar mass.
Through the use of two-dimensional hydrodynamic simulations, \citet{baraffe23} quantifies the strength of overshooting in \gls*{zams} stars and indeed finds an increase of overshooting distance with stellar mass.
Despite this, the increase in overshooting distance is not enough to resolve the discrepancy between models and observations alone.
Furthermore, \citet{morison24} found that the stratification of helium abundance atop the cores of evolved $5M_\odot$ \gls*{ms} stars limits the overshooting in the radiative envelope of their two-dimensional hydrodynamic simulations.
Using current 1D stellar evolution models, \citet{johnston24} derive predictions for penetration distance above the convective core as a function of mass, and conclude that additional mixing processes are required to match their predictions with inferences from observations, even for less massive stars with $M \lesssim 5M_\odot$.
Hence, we now turn to \glspl*{igw} as another mechanism which could contribute to the mixing in stable radiative zones.

While also a well-known process in oceans and planetary atmospheres \citep{lighthill78}, vertical wave mixing in those cases is usually discussed in the context of non-linear effects and wave breaking \citep{whalen2020}.
In contrast, \glspl*{igw} in stars are typically linear \citep{press81} and characterising the resulting mixing remains difficult.
There have been many efforts to quantify stellar mixing due to \glspl*{igw}, including both theoretical approaches \citep[e.g.][]{press81,lopez91,schatzman96,montalban00,talon05} and more recently numerical simulations \citep[e.g.][]{rogers17,higl21,varghese23,herwig23}.

\citet{press81} consolidated the linear theory of propagating \glspl*{igw} in the stellar context, including the effects of thermal diffusion.
From this, an estimate for mixing is derived based on the entropy lost by waves to the surroundings.
\citet{lopez91} proposed a mechanism by which the shearing motions of the waves induce weak localised flows via the Kelvin-Helmholtz instability.
\citet{schatzman96} and the series of papers ending with \citet{montalban00} build on the work of \citet{press81} to explain the relation between this Li depletion and stellar properties such as mass, age, and rotation.
\citet{talon05} incorporated effects of \glspl*{igw} alongside rotation into stellar evolution, though neglect convective overshooting and penetration.
All the aforementioned analytical models are based on assumptions of the spectrum of \glspl*{igw} generated at the convective-radiative interface.
Thus, we are motivated to test a selection of these models with the \gls*{igw} spectra obtained from our hydrodynamic simulations.
Following an outline of the numerical simulations in \secref{sec:models}, we present an analysis of the wave spectra in \secref{sec:waves} and use these to evaluate the \gls*{igw} mixing models of \citet{press81} and \citet{lopez91} in \secref{sec:wave-mixing}.

On the numerical front, \citet{rogers17} performed two-dimensional spherical hydrodynamic simulations of a $3M_\odot$ stellar model on the \gls*{ms}, and used tracer particles embedded in the flow to track the movement of material.
The work suggested that radial transport of material by \glspl*{igw} can be modelled as a diffusive process, and derived large diffusion coefficients proportional to the amplitude of the waves throughout the radiative zone.
\citet{higl21} and \citet{varghese23} used similar methods, though \citet{varghese23} finds diffusion coefficients approximately three orders of magnitude below those of \citet{rogers17} for the same mass star.
\citet{higl21} comments on the difficulty of numerically integrating the tracer particle trajectories, so does not track them over large time periods.

In \secref{sec:tracers}, we apply the tracer particles method of \citet{rogers17} to our simulations of massive stars, and discuss the results together with the analytical predictions of \citet{press81} and \citet{lopez91}.
None of the aforementioned studies implementing this method track the tracers over time periods comparable to the timescales of the diffusion coefficients which are measured (see \secref{sec:particle-method} for details), due to numerical artefacts which emerge over time or otherwise.
The difficulty in reducing these numerical artefacts will be discussed in \secref{sec:tracers}, since understanding them is imperative when measuring small diffusion coefficients acting over long time periods.
Despite this limitation, the diffusion coefficient profile obtained by \citet{rogers17} for stable radiative regions has been implemented in one-dimensional stellar evolution models for asteroseismic studies \citep[e.g.][]{pedersen2018,michielsen2021,michielsen2023} as a diffusive mixing profile.
For this reason, it is important to compare the mixing profiles of previous works with those computed using differing numerical approaches, and to draw links to theoretical models.
More rigorous diffusion profiles will provide confidence in the results of stellar evolution models and the asteroseismic studies that utilise them.

%% file: src/02-models.tex

\section{Stellar models and numerical simulations}\label{sec:models}

\subsection{Initial 1D stellar models}
\begin{figure}
	\centering
	\includegraphics[width=\linewidth]{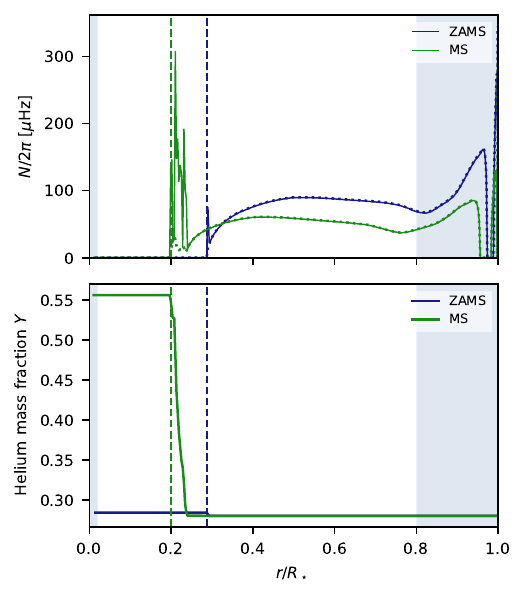}
	\caption{
		Radial profiles of properties of the 1D stellar evolution models in \tabref{tab:models}.
		\emph{Top}: The Brunt-V\"ais\"al\"a frequency (with thermal contributions as dotted lines), and \emph{bottom}: the helium mass fraction $Y$.
		The greyed out areas indicate those which are not simulated in MUSIC.
		Vertical dashed lines indicate the convective core boundary of each of the models, according to the Schwarzschild criterion.
	}
	\label{fig:model-profiles}
\end{figure}

We perform two-dimensional (2D) numerical simulations of the convective core of a non-rotating 20 $\msol$ star with an initial helium abundance in mass fraction $Y$=0.28 and metallicity $Z$=0.02 at two different stages of evolution, namely on the \gls*{zams} and later on the \gls*{ms}.
The initial stellar models are computed  with the one-dimensional Lyon stellar evolution code \citep{baraffe91, baraffe98}, using the same opacities and equation of state as MUSIC. 
The ZAMS  model is the same as in \citet{baraffe23} with a central abundance of helium $Y_{\rm c}$=0.2838 (\figref{fig:model-profiles}),  {\it i.e.} only $\sim$ 1\% of the central hydrogen has been depleted.
The \gls*{ms} model has a central helium mass fraction $Y_{\rm c}$=0.5565.
The 1D stellar models rely on the Schwarzschild criterion for the onset of convective instability and do not account for overshooting at the convective core boundary.
These assumptions to construct the initial models will be discussed in \secref{sec:discussion}.
The properties of the stellar models are provided in \tabref{tab:models}.

\subsection{2D hydrodynamical models}
The 2D numerical simulations are performed with the fully compressible time-implicit code MUSIC.
A description of MUSIC and of the time-implicit integration can be found in \citet{viallet11, viallet16, goffrey17}.
Since this work uses the same numerical setup as that of \citet{baraffe23}, we direct the reader there for the description of the equations which are solved (see their Section 2).
We only remind that the 2D simulations are performed in a spherical shell using spherical coordinates with angular extent from $\theta = 0 ^\circ$ to $\theta = 180 ^\circ$ and assuming azimuthal symmetry in the $\phi$-direction.
The inner radius $r_{\rm in}$ is defined at 0.02 $R_\star$ and the outer radius $r_{\rm out}$ at 0.8 $R_\star$, increasing further the radial extension of the 20 $\msol$ star model studied in \citet{baraffe23}.
The grid has uniform spacing in the r and $\theta$ coordinates.
The radial resolution is set such that there are radially 140 grid cells per pressure scale height at the convective core boundary --- see the aforementioned study for a justification of this choice.
The boundary conditions are the same as in \citet{baraffe23} (see their Sec. 2.2).
In particular, for the velocity, we impose reflective conditions at the radial and polar boundaries, corresponding to
\begin{itemize} 
\item{}  $\vel_r$ = 0 and ${\partial \vel_\theta \over \partial r}=0$ at $r_{\rm in}$ and $r_{\rm out}$,
\item{}  ${\partial \vel_r \over \partial \theta}=0$ and  $\vel_\theta$ = 0 at $\theta=0^\circ$ and $\theta=180^\circ$,
\end{itemize}
\noindent with $v_r$ and $v_{\theta}$  the radial and angular velocities, respectively.

\begin{table}
	\caption{Properties of the initial stellar models for a 20 $\msol$ star used for the 2D hydrodynamical simulations: stellar luminosity, stellar radius, mass and radius of the convective core (corresponding to the location of the Schwarzschild boundary) and central helium mass fraction $Y$. }
	\label{tab:models}
	\centering
	\begin{tabular}{l l l l l l }
		\hline \hline
		Model &  $L_\star/L_\odot^{a}$ & $R_\star$ (cm) &  $M_{\rm conv}/\msol$ &  $r_{\rm conv}/R_\star$ & $Y_{\rm c}$ \\
		\hline
		ZAMS & 4.29$\times$10$^{4}$ & 4.0172$\times$10$^{11}$ & 8.79 & 0.287 & 0.2838 \\
		MS &  6.12$\times$10$^{4}$ & 5.5877$\times$10$^{11}$ & 7.07 & 0.196 &  0.5565\\
		\hline
		\multicolumn{3}{l}{$^a$ We use $L_\odot = 3.839 \times 10^{33}$ erg/s.}
	\end{tabular}
\end{table}

Reflective boundary conditions at the upper radial boundary can enhance the amplitude of standing waves formed in the stably stratified region above the convective core.
The top boundary will artificially reflect propagating waves that would be damped in the region between $r_{\rm out}$ and $R_\star$, contributing to the formation of standing waves.
Even modes which would not be entirely damped in the real star will be strengthened by the reduced cavity size, as it shortens the distance over which waves must propagate to form them.
Hence, we will take steps to assess the impact of the outer boundary condition on the waves that arise, ensuring that it does not alter conclusions we make in this work.
To this end, we limit wave reflection from the outer radial boundary by running simulations as in \citet{couston18, lecoanet21}, adding a damping term, $-\rho \velvec \, h(r) / \tau_{\rm damp}$, in the momentum equation.
This yields
\begin{eqnarray}\label{eq:music-momentum}
	\frac{\partial \rho \velvec}{\partial t} &=& - \vec \nabla \cdot (\rho \velvec \otimes \velvec)-\vec \nabla p + \rho \vec g -\rho \velvec \frac{h(r)}{\tau_{\rm damp}}
\end{eqnarray}
\noindent where $\rho$ is the density,  $\vec \vel$ the velocity, $p$ the gas pressure, and $\vec g$ the spherically symmetric gravitational acceleration \citep[see][]{baraffe23}.
$h(r)$ is the spatial profile of the damping layer, defined as
\begin{eqnarray}
	h(r) = {1 \over 2} (1 - \cos (\pi X)), \\
	X={r-r_0 \over r_1 - r_0},         \label{eq:r0r1}
\end{eqnarray}
\noindent so that waves are damped in the region $r > r_0$ and $h(r)$ varies from 0 to 1 between $r_0$ and $r_1$.
Such damping layers will most strongly damp waves whose radial wavelength is sufficiently smaller than the width of the layer, and whose group velocity is slow enough that a given wavefront spends approximately a damping timescale or larger within the damping layer.
Hence, the damping timescale $\tau_{\rm damp}$ is chosen such that a significant proportion of the most powerful waves are damped, preventing them from forming standing modes.
The simulations performed with damping layers in this work are listed in \tabref{tab:damped-simulations}.

\begin{table}
	\caption{
		Damping layer simulations performed in this work.
		All simulations in this table have the same properties as their listed reference model, aside from the given damping parameters.
	}
	\label{tab:damped-simulations}
	\centering
	\begin{tabular}{lccc}
		\hline \hline
		Reference Model & $\tau_{\rm damp}$ (s) & $r_0/R_\star$ & $r_1/R_\star$ \\
		\hline
		ZAMS & $1.8\times 10^3$ & 0.64 & 0.80 \\
		ZAMS & $1.0\times 10^5$ & 0.70 & 0.80 \\
		ZAMS & $1.0\times 10^5$ & 0.78 & 0.80 \\
		MS & $2.6\times 10^3$ & 0.64 & 0.80 \\
		MS & $1.0\times 10^5$ & 0.70 & 0.80 \\
		MS & $1.0\times 10^5$ & 0.78 & 0.80 \\
		\hline
	\end{tabular}
\end{table}

We have also performed simulations for the ZAMS model with artificial enhancement of the stellar luminosity 
and the thermal diffusivity by a factor of 10$^3$, a common practice in stellar hydrodynamics simulations \citep[][]{rogers06, meakin07, tian09, brun11, rogers13, brun17, hotta17, cristini17, edelmann19, horst20}.
In previous works, we have shown that this modification can have a significant impact on the properties of the convective penetration layer  \citep{baraffe21, baraffe23} and  on  the amplitude and frequency range of \glspl*{igw} \citep{lesaux22, lesaux23}.
The boosted ZAMS model thus provides an interesting experiment to test whether increased luminosity can enhance the process of wave mixing, since the amplitude of the waves increases with luminosity, but radiative damping is strengthened with larger thermal diffusivities.
This model can also provide a test for different theoretical models for wave mixing, the efficiency of which depends on the wave frequency (see \secref{sec:wave-mixing}).

\subsection{Properties of the numerical simulations}

The properties of all simulations are summarised in \tabref{tab:simulations}.
We define $t_{\rm steady}$ as the time required to reach a steady state for convection, characterised by the total kinetic energy of the system reaching a plateau.
At $t_{\rm steady}$, the value of the kinetic energy starts to stabilise and from this time it remains roughly constant with time.
The simulations are stopped at time $t_{\rm sim}$ provided in \tabref{tab:simulations}.
A global convective turnover time $\tau_{\rm conv}$ is estimated based on the RMS velocity $ \vel_\mathrm{RMS}(r,t)$ at radius $r$ and time $t$, which characterises a bulk convective velocity.
We define $\tau_{\rm conv}$ by
\begin{equation}
	\tau_{\rm conv} =  \Big \langle {{ \int_{r_{\rm in}}^{r_{\rm conv}} {{\rm d}r \over \vel_\mathrm{RMS}(r,t)  }  }  }\Big \rangle_t,
	\label{eq:tconv}
\end{equation}
where $r_{\rm conv}$ is the position of the convective-radiative boundary according to the Schwarzschild criterion, and the RMS velocity is given by
\begin{equation}\label{eq:vrms}
	\vel_\mathrm{RMS}(r,t) = \sqrt {\langle \vel^2(r,\theta,t) \rangle_\theta},
\end{equation}
with $\vel^2 = \vel_r^2 + \vel_{\theta}^2$.
Time averages are denoted by $\langle \rangle_t$ and calculated between $t_{\rm steady}$ and $t_{\rm sim}$.
For any quantity $f$ we define:
\begin{equation}
	\big \langle  f \big \rangle_t = {1 \over (t_{\rm sim} - t_{\rm steady})} \int_{t_{\rm steady}}^{t_{\rm sim}} f {\rm d}t
\end{equation} 
 The volume-weighted average in the angular direction $\langle \rangle_\theta$ is defined for any quantity $f$ as:
\begin{equation}
	\label{horizontal}
	\big \langle f(r,\theta,t) \big \rangle_\theta = {\int_\theta  f(r,\theta,t) {\rm d}V(r,\theta) \over \int_\theta {\rm d}V(r,\theta)}.
\end{equation}
\noindent

\begin{figure}
	\centering
	\begin{subfigure}{\columnwidth}
		\includegraphics[width=\linewidth]{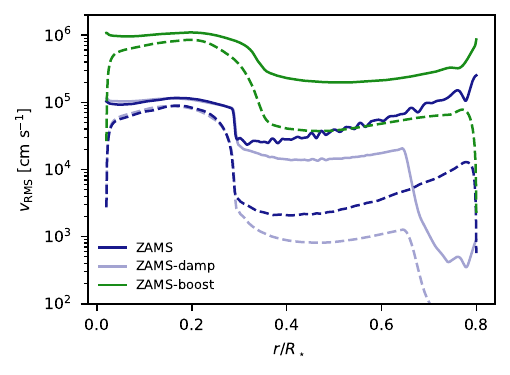}
		\caption{Simulations initialised from the ZAMS 1D model}\label{fig:vrms-ZAMS}
	\end{subfigure}
	\begin{subfigure}{\columnwidth}
		\includegraphics[width=\linewidth]{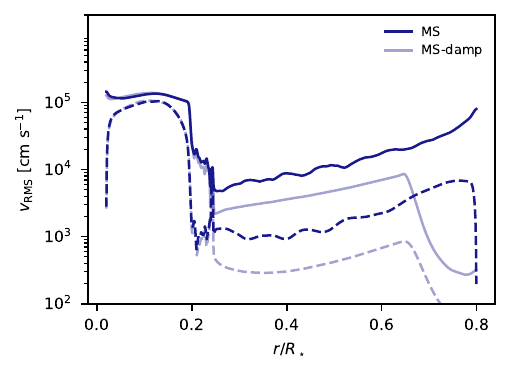}
		\caption{Simulations initialised from the evolved MS 1D model}\label{fig:vrms-MS}
	\end{subfigure}
	\caption{
		Total and radial $v_{\rm RMS}$ profiles as solid and dashed lines respectively, averaged in time and horizontally as per \eqref{eq:vrms}.
		The simulations with damping layers shown here are those with the smallest $\tau_{\rm damp}$ in \tabref{tab:damped-simulations}, with $r_0 = 0.64R_\star$.
	}
	\label{fig:vrms-profiles}
\end{figure}

\figref{fig:vrms-profiles} shows the profiles of RMS total velocity and RMS radial velocity for all simulations.
An unexpected feature of these profiles is the peak in radial velocities at $r \sim 0.25R_\star$ of the evolved \gls*{ms} simulations seen in \figref{fig:vrms-MS}.
The location of this peak corresponds to the top of the chemical gradient atop the convective core (shown in \figref{fig:model-profiles}), and is due to a thin unstable layer that arises here in the MUSIC simulations, but is not present in the 1D stellar evolution model.
We will discuss this behaviour further in \secref{sec:power-spec-ms}, making the argument that it does not significantly affect our results.

\begin{table*}
	\caption{Main properties of the 2D simulations.}
	\label{tab:simulations}
	\begin{tabular}{lcccccc}
		\hline \hline
		Model & $L$ (erg/s) & $N_r \times N_\theta$ &  $\tau_{\rm conv}^{a}$ (s) &  $N_{\rm conv}^{b}$ &  $t_{\rm steady}^{c}$ (s) & $t_{\rm sim}^{d}$ (s) \\
		\hline
		ZAMS  & 1.649 $\times  10^{38}$  & 1344x896  & 1.2 $\times 10^6$ & 134 & 9.00 $\times 10^7$ & 2.49 $\times 10^8$ \\
		ZAMS-boost & 1.649 $\times  10^{41} $  & 1344x896   & 1.2 $\times 10^5$ & 166  & 1.96 $\times 10^7$ & 3.97 $\times 10^7$ \\
		MS &  2.351 $\times  10^{38}$ & 1824x832  & 9.0 $\times 10^5$ & 204 & 1.52  $\times 10^8$ &   3.33 $\times 10^8$ \\
		\hline

		\multicolumn{7}{l}{$^a$ Convective turnover time (see \eqref{eq:tconv}).} \\
		\multicolumn{7}{l}{$^b$ Number of convective turnover times covered by the simulation once  steady state convection is reached.}\\
		\multicolumn{7}{l}{$^c$ Physical time to reach a  steady state for convection.}\\
		\multicolumn{7}{l}{$^d$ Total physical runtime of the simulation.} \\
	\end{tabular}
\end{table*}

%% file: src/03-waves.tex

\section{Wave Analysis}\label{sec:waves}

\subsection{Power spectra}

\begin{figure*}
	\centering
	\includegraphics[]{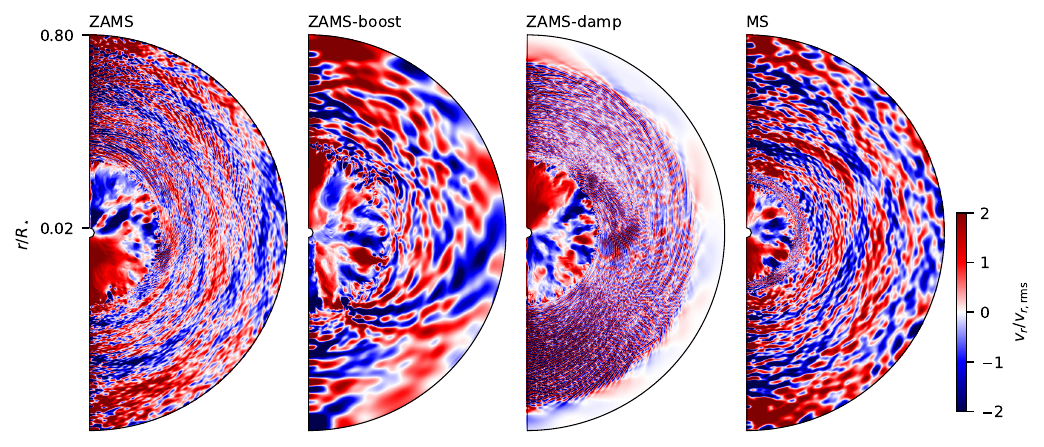}
	\caption{
		Snapshots in time of the radial velocity $v_r(r, \theta)$, scaled by the RMS profile $v_{r, \rm RMS}(r)$.
		The damping layer simulation shown here is based on the ZAMS model, and uses $\tau_{\rm damp} = 1.8 \times 10^3$ (see \tabref{tab:damped-simulations}).
	}
	\label{fig:snapshots}
\end{figure*}

Internal gravity waves (\glspl*{igw}) propagate according to the dispersion relation
\begin{equation}\label{eq:igw-dispersion}
\omega^2 = N^2 \frac{k_h^2}{k^2} = N^2 \cos^2\alpha,
\end{equation}
where $\omega$ is the wave's angular frequency in $\si{\radian\per\second}$, $k = |\vec{k}| = |\vec{k_r} + \vec{k_h}|$ is the wavenumber, with $k_r$ and $k_h = \frac{1}{r}\sqrt{\ell(\ell + 1)}$ as the radial and horizontal (angular) components respectively \citep{lighthill78}.
Since the frequency of the wave depends only on the orientation of the wave vector, we can write the dispersion relation in terms of $\alpha$, the angle between the wave vector and the horizontal direction.
The Brunt-V\"ais\"al\"a (or buoyancy) frequency $N$ sets the maximum oscillation frequency of the waves, and is given (in $\si{\radian\per\second}$) by
\begin{equation}\label{eq:bvf}
N^2 = g \left( \frac{1}{\Gamma_1} \frac{d\ln p}{dr} - \frac{d\ln \rho}{dr} \right)
\end{equation}
where
\begin{equation}
\Gamma_1 = \left( \frac{d\ln p}{d\ln \rho} \right)_{\rm ad}
\end{equation}
is the first adiabatic exponent.
The radial profiles of $N$ are shown for each of the two 1D stellar models in \figref{fig:model-profiles}.
$N^2 > 0$ implies a stably stratified region such as the radiative envelope in each of the models, whereas an unstable region such as the convective core will have $N^2 \le 0$.
\glspl*{igw} only propagate in stably stratified regions where buoyancy is strong enough to act as a restoring force for the fluid, and the waves will become evanescent upon entering an unstable region.
Hence, we see the waves confined to the radiative regions of the simulations, as is visible in the snapshots of normalised radial velocities shown in \figref{fig:snapshots}.

\begin{figure*}
	\includegraphics[width=\linewidth]{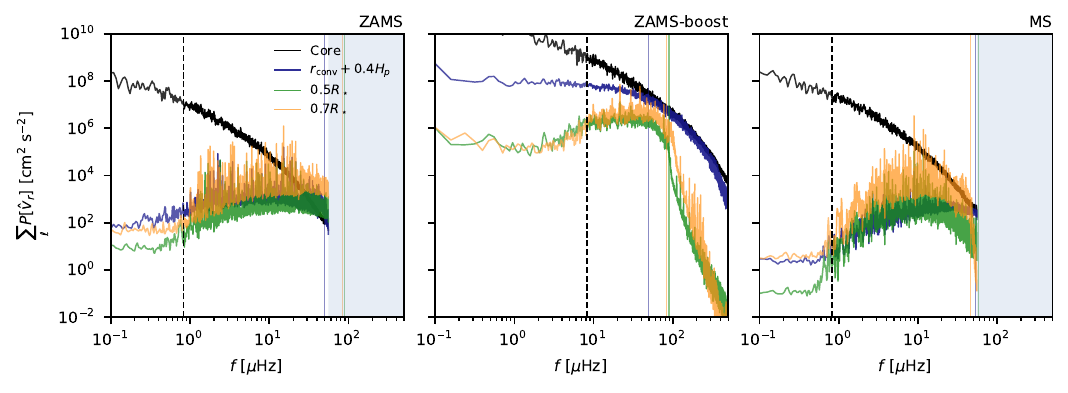}
	\caption{
		Power spectra for three simulations, summed over all values of $\ell$ and taken at four different radii --- one pressure scale height into the convective core, $0.4$ pressure scale heights above it (within the $N^2$ peak of the \gls*{ms} model), approximately half way through the simulated radiative envelope, and $0.1R_\star$ from the outer boundary of the simulation.
		The convective frequency is indicated with a vertical dashed line in each figure.
		The Brunt-V\"ais\"al\"a frequency at each radius is indicated as a solid vertical line of the corresponding colour.
		The shaded regions indicate frequencies higher than the Nyquist frequency of the simulations' outputs used to compute the power spectra.
	}
	\label{fig:power-spec}
\end{figure*}

To analyse and compare the waves in each simulation, we computed the power spectrum of the radial velocity $P[\hat{v}_r]\left(\omega, r, \ell\right)$ by transforming the radial velocity field of the simulation $v_r(t, r, \theta)$ into spherical harmonics $\hat{v}_r(t, r, \ell)$, and then computing the Fourier transform of the time series.
Details of the methods used for this are given in \cite{lesaux22}.
Some power spectra of the simulations performed in this work are shown in \figref{fig:power-spec}, after taking a horizontal average by summing over the spherical harmonics.

In our simulations, we find strong standing g-modes which can be characterised by a single pair of $\ell$ and $\omega$, hence appearing as sharp peaks in \figref{fig:power-spec}.
As the waves are excited by the core convection at the convective-radiative boundary, we observe a fall off in power at frequencies approximately below the convective frequency $f_{\rm conv} = \tau_{\rm conv}^{-1}$ (\tabref{tab:simulations}) of the simulation.
As there are no mechanisms to excite \glspl*{igw} at these low frequencies, they cannot be reliably considered as signatures of wave motion \citep{lecoanet13} which will be important to note when considering the contribution of \glspl*{igw} to mixing in \secref{sec:wave-mixing}.
Consistent with this, we do not see significant g-modes with frequencies below the convective frequency of each simulation.

\subsection{Comparison with linear theory}\label{sec:linear-theory}

\cite{press81,kumar99} derive an expression for the amplitude of an \gls*{igw} propagating linearly through a stably stratified region as
\begin{equation}\label{eq:linear-amplitude}
u(r, \ell, \omega) = u(r_e, \ell, \omega) {\left(\frac{\rho}{\rho_e}\right)}^{-1/2} {\left(\frac{r}{r_e}\right)}^{-3/2} {\left(\frac{N^2 - \omega^2}{N_e^2 - \omega^2}\right)}^{-1/4} e^{-\tau/2}
\end{equation}
where the subscript $e$ implies quantities evaluated at the radius of excitation $r=r_e$, and
\begin{equation}\label{eq:linear-damping}
\tau = \tau(r, \ell, \omega) = {[\ell(\ell + 1)]}^{3/2} \int_{r_e}^r \kappa_{\rm rad} \frac{N_t^2 N}{\omega^4} {\left(\frac{N^2}{N^2 - \omega^2}\right)}^{1/2} \frac{dr}{r^3}
\end{equation}
represents the strength of radiative damping of the wave \citep{zahn97}, where $N_t$ is the thermal contribution to the buoyancy frequency (shown in \figref{fig:model-profiles}) expressed as
\begin{equation}
N_t^2 = -\left.\frac{\partial\ln\rho}{\partial\ln T}\right|_{p, \mu} \frac{g}{H_p} \left(\nabla_{\rm ad} - \nabla\right),
\end{equation}
where $\nabla \equiv \partial\ln T / \partial \ln p$.
\cite{press81} derives \eqref{eq:linear-amplitude} from the linearised anelastic fluid equations with non-adiabatic effects, obtaining a solution using the WKB approximation \citep[see, for example][Ch. 8]{Schiff55}, and \citet{zahn97} adapts this to include gradients in molecular weight $\mu$ (significant for the \gls*{ms} model in this work).
Approximating the perturbations as linear holds for waves with small amplitudes, so we expect to see a departure from this theory for \glspl*{igw} with a significant amount of power, particularly at radii close to the outer boundary of the simulation.
Radial velocities there can be increased by up to an order of magnitude compared to velocities at the convective-radiative boundary (see \figref{fig:vrms-profiles}) due to the large decrease in density towards the outer boundary, which amplifies the waves and gives rise to the increase in power from $0.5R_\star$ to $0.7R_\star$ that is visible in all panels of \figref{fig:power-spec}.

\begin{figure*}
	\includegraphics[width=\linewidth]{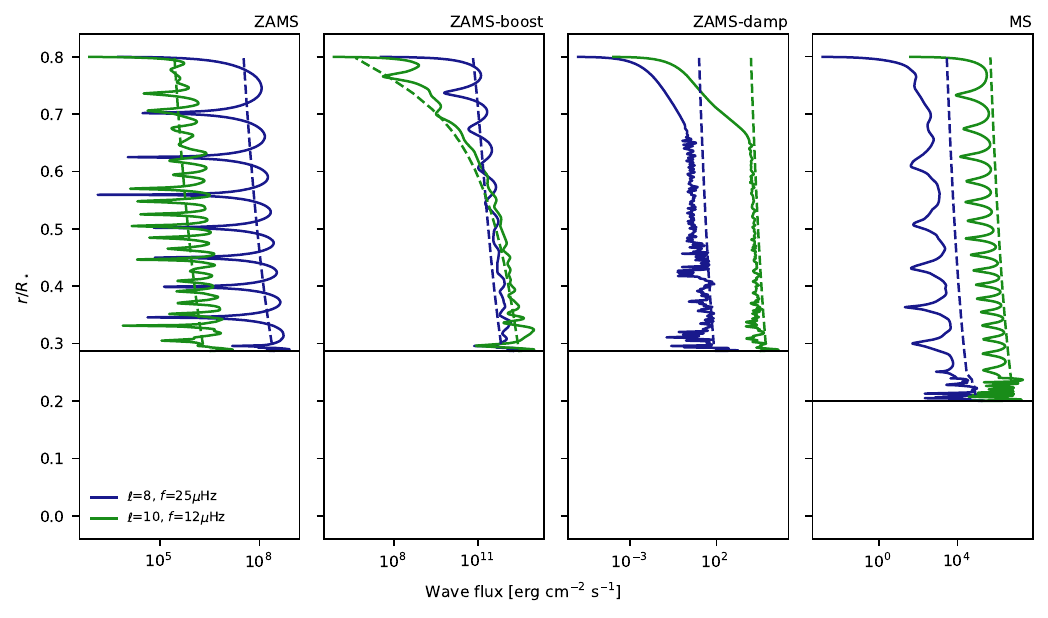}
	\caption{
		Radial wave fluxes from the numerical simulations (solid lines) for two selected modes, computed from the radial velocity power spectrum as per \eqref{eq:fwave}.
		The dashed lines show equivalent fluxes computed using the wave amplitudes from linear theory (\eqref{eq:linear-amplitude}), and the horizontal line shows the convective-radiative boundary.
		The damping layer simulation shown here is based on the ZAMS model, and uses $\tau_{\rm damp} = 1.8 \times 10^3$ (see \tabref{tab:damped-simulations}).
	}
	\label{fig:wave-flux}
\end{figure*}

\eqref{eq:linear-amplitude} predicts that high frequency waves with low angular degree $\ell$ should be damped the least.
This effect is evident in the power spectra (\figref{fig:power-spec}) where the distribution of power at $0.7R_\star$ is slightly shifted towards higher frequencies of $f \gtrsim 10 f_{\rm conv}$ compared with the spectrum at $0.5R_\star$.
To further compare the waves with linear theory, we compute their fluxes by taking the kinetic energy density of a wave $\frac{1}{2} \rho \langle u^2 \rangle$, taking twice this to obtain the total energy density (according to the equipartition of energy), and multiplying by the radial group velocity \citep{zahn97} to obtain
\begin{equation}\label{eq:fwave-vrms}
F_{\rm wave} \equiv \rho \vel_{\rm rms}^2 \cdot u_{g, r}
\end{equation}
where $u_{g, r}$ is the radial group velocity.
Substituting in the power spectrum gives the flux of each \gls*{igw} as
\begin{equation}\label{eq:fwave}
F_{\rm wave}(r, \ell, \omega) = \rho \frac{\sqrt{N^2 - \omega^2}}{k_h} P[\hat{\vel}_r](r, \ell, \omega).
\end{equation}
See \citet{morison24} for a more detailed version of this derivation.
\figref{fig:wave-flux} shows $F_{\rm wave}(r, \ell, \omega)$ computed from the simulations' power spectra for two different modes, alongside the predictions from linear theory using the theoretical wave amplitudes from \eqref{eq:linear-amplitude}.
The formation of g-modes is evident from periodic oscillations in the fluxes, and we find that the envelope of the oscillations matches well with linear theory in most cases.
The evolved \gls*{ms} simulations exhibit a slight discrepancy between their fluxes and the prediction of linear theory, largely due to the aforementioned thin unstable layer.
Also shown are the fluxes of two modes in a simulation with a damping layer (discussed in \secref{sec:discussion}), which display g-modes with greatly reduced amplitudes compared with their non-damped counterparts, which confirms that the wave reflections at the numerical outer boundary are significantly reduced.

\subsection{Effect of boosting}
For this work, we performed one \emph{boosted} simulation with luminosities and thermal diffusivities artificially enhanced by a factor of $10^3$.
The boosting not only increases the overall power in the waves, but also has a significant effect on the spectrum of \glspl*{igw}, similarly observed in \citet{lesaux23} for a $5M_\odot$ \gls*{zams} model.
We consider this boosted model not for its physical relevance, but because enhancing luminosity is a common technique used to reduce the thermal timescale of numerical simulations and it is informative to see what effect, if any, this should have on mixing.
Even without considering power spectra, the difference is evident in the radial velocity fields in \figref{fig:snapshots}, with the boosted model giving rise to larger wavelengths and hence visually larger structures.
Boosting the model decreases the convective turnover timescale by a factor of ten, which is visible in \figref{fig:power-spec} as a peak at higher frequencies and a lack of g-modes below $\sim 10 \si{\micro\hertz}$.
The buoyancy frequency is not altered by the artificial boosting however, so the increased Nyquist frequency of the simulation by a factor of ten  reveals the steep fall-off in power at $\omega = N\sim 100 \si{\micro\hertz}$.
The spectrum just above the core strongly resembles that of the convection itself, which is consistent with the enlarged penetration layers that develop in the highest luminosity models of \citet{baraffe23}.

Since the radiative diffusivity is enhanced, the radiative damping predicted by linear theory is strengthened by the same factor, as per \eqref{eq:linear-damping}.
This agrees with the wave fluxes in \figref{fig:wave-flux}, with the higher angular degree $\ell$ and lower frequency $\omega$ mode decaying towards the surface, an effect not observed in the non-boosted simulations for the same mode.

\subsection{The compositional gradient in the main-sequence model}\label{sec:power-spec-ms}
The mid \gls*{ms} model in \tabref{tab:models} has a gradient in the helium mass fraction $Y$ just above the convective core, a significant feature not present in the ZAMS model where the boundary only has a small step in helium abundance.
This gradient in composition feeds into the buoyancy frequency $N$ through the density gradients present in \eqref{eq:bvf}, and results in a large peak in $N^2$ over a region spanning about half a pressure scale height (see \figref{fig:model-profiles}).
As discussed in \citet{morison24}, the assumptions used in the 1D stellar evolution models will affect the properties of this peak in $N^2$, but its existence is guaranteed so long as the core retracts and leaves behind a chemical gradient \citep{miglio2008}.
Hence, the qualitative analysis surrounding the layer is relevant no matter the 1D model employed.

Recalling that the buoyancy frequency defines the maximum frequency of oscillation of \glspl*{igw}, we can predict the appearance of some high frequency waves in this region.
We indeed observe this effect in \figref{fig:power-spec}, with the spectrum within the $N^2$ peak showing substantially more power at high frequencies $\omega \gtrsim 20\si{\micro\hertz}$ compared to spectra above the chemical gradient.
Due to the buoyancy frequency defining the turning point of \glspl*{igw} of equal frequency, the subsequent dip in $N$ just above the peak confines these high frequency \glspl*{igw} to this small region, leading to their decay at larger radii.

A sharp narrow peak in radial velocity atop the $N^2$ peak is visible in \figref{fig:vrms-MS}, due to a convectively unstable layer that arises in the MUSIC simulations as a result of slight changes in the stratification while the simulation relaxes into a steady state after being initialised with the 1D model in \tabref{tab:models}.
However, we argue that this layer does not significantly affect the wave spectra, since the width of the layer is so thin (on the order of a few MUSIC grid cells) that only waves with a similarly short radial wavelength are notably affected.
These waves have very low frequencies $f < f_{\rm conv}$ (\eqref{eq:igw-dispersion}), little power (\figref{fig:power-spec}), and are possibly not well resolved in the simulations due to the short wavelengths.
The \glspl*{igw} of interest $f \gtrsim f_{\rm conv}$ are largely unaffected by this feature, as seen by the match between wave fluxes and linear theory (to within a scaling factor) in \figref{fig:wave-flux}.

The waves have less power overall than the ZAMS simulation, leading to lower RMS velocities in the radiative zone despite the cores of the two models being similar in this regard.
This agrees with the results of the $5M_\odot$ model in \citet{morison24}, which finds that waves are heavily damped in the $N^2$ peak region above the core.
\citet{morison24} finds that the peak in $N^2$ also limits convective penetration distance in evolved models, due to the hindered radial motion.
In \figref{fig:power-spec}, the ZAMS simulation shows an increase in the power of low frequencies $\omega \lesssim 10\si{\micro\hertz}$ at a pressure scale height above the core, frequencies which are characteristic of the core convection and the associated penetrative plumes.
This effect is much less pronounced in the evolved MS simulation, supporting the results of \citet{morison24}.

%% file: src/04-mixing.tex

\section{Wave mixing mechanisms and diffusion coefficients}\label{sec:wave-mixing}

\subsection{Wave mixing models}\label{sec:mixing-models}
In this work, we will explore two possible mechanisms described in the literature to drive mixing by \glspl*{igw}.
The first is directly linked to the effect of thermal diffusion \citep{press81} and the second is 
due to weakly turbulent flows induced by local vertical shear produced by the waves \citep[][]{lopez91}.
\citet{press81} describes a mechanism by which thermal diffusion produces a loss of entropy and introduces an irreversible process during the oscillatory motion of fluid elements.
These irreversible effects generate an RMS displacement of each fluid element, because they reach a position after an oscillation period which is slightly shifted from their starting point due to the loss of energy by thermal diffusion. This results in a
macroscopic diffusive process which can transport chemical elements.
By applying linear theory, \citet{press81} obtains, to an order of magnitude, a diffusion coefficient for a monochromatic \gls*{igw} with frequency $\omega$, vertical wave number $k_{\rm v}$ and vertical velocity $u_{\rm v}$:
\begin{equation}
D_{P81} = {\epsilon^4 \kappa_{\rm T}^2 k_{\rm v}^2 \over \omega},
\end{equation}
\noindent with $\kappa_{\rm T}$ the thermal diffusivity and $\epsilon$ the nonlinearity parameter defined by
\begin{equation}\label{eq:non-lin-param}
\epsilon \equiv {k_{\rm v} u_{\rm v} \over \omega}.
\end{equation}

In the stellar models considered in this work, the thermal diffusivity is set to the radiative diffusivity $\kappa_{\rm rad}$ given by
\begin{equation}
\kappa_{\rm T} = \kappa_{\rm rad} = \frac{16 \sigma T^3}{3\kappa \rho^2 c_P}
\end{equation}
\noindent where $\kappa$ is the Rosseland mean opacity, $\sigma$ the Stefan-Boltzmann constant and $c_P$ the specific heat capacity at constant pressure. The diffusion coefficient from \citet{press81} thus writes:
\begin{equation}\label{eq:diffusion-p81}
D_{P81} = { \kappa_{\rm rad}^2 k_{\rm v}^6 u_{\rm v}^4 \over \omega^5}.
\end{equation}
The condition for applying linear theory is $\epsilon < 1$.
Further models relying on this mechanism were developed, in particular by \citet[][]{schatzman93, montalban94, schatzman96, montalban00}, to derive a total diffusion coefficient assuming some power spectrum for the ensemble of waves generated at the convective boundary. 

\citet[][]{lopez91} describes a different mechanism resulting from the property that \glspl*{igw} (standing and progressive) with frequencies $ \omega \ll N$ oscillate in nearly horizontal directions, as implied by their dispersion relation (see \eqref{eq:igw-dispersion}).
They can thus produce horizontal velocity shears which, in the presence of thermal diffusion, result in small-scale, weakly turbulent flows that can induce mixing.
This follows arguments developed by \citet[][]{townsend58} and \citet[][]{zahn83} that thermal diffusion acting on turbulent elements in an unstable shear flow will result in a less strict condition than the usual Richardson condition to produce turbulence via a Kelvin-Helmholtz instability.
Thermal diffusion indeed reduces temperature fluctuations and thus buoyancy effects.
If a turbulent element has a typical size $l$ and velocity $v$ such that its cooling time $\tau_{\rm cool} \sim l^2/\kappa_{\rm T}$ is smaller than its overturning time $\tau_{\rm ov} \sim l/v$, it will feel a buoyancy force reduced by a factor (1 + $\tau_{\rm cool} / \tau_{\rm ov}$) \citep[][]{zahn83}.
This results in a less stringent Richardson criterion for instability, thus limiting the stabilising effect of the stratification.
Applying these arguments, \citet[][]{lopez91} obtain a diffusion coefficient for a given wave with frequency $\omega$ and angular degree $\ell$, (for $ \omega \ll N$):
\begin{equation}\label{eq:diffusion-gls91}
D_{GLS91} = { \left ( \ell (\ell + 1) \right )^{3/2} \over 4 \rho r^3} {N \over \omega^4} \kappa_{\rm rad} F_{\rm wave} (\omega,\ell),
\end{equation}
with $F_{\rm wave}$ the wave energy flux, directly provided by the numerical simulations via \eqref{eq:fwave}.
\citet{lopez91} propose a condition for producing mixing via this mechanism, with the assumption that the flow in an \gls*{igw} of frequency $\omega$ must be steady enough to produce small scale turbulence via the horizontal shear.
This is possible if the shear rate $k_{\rm v} u_{\rm h}$ is larger than the wave frequency. The condition for mixing suggested by \citet[][]{lopez91} is thus:
\begin{equation}\label{eq:criterion-gls91}
k_{\rm v} u_{\rm h} > \omega.
\end{equation}
We will discuss this assumption in the context of the simulations in \secref{sec:discussion}.

\subsection{Maps of diffusion coefficients from the numerical simulations}\label{sec:diffusion-maps}
\begin{figure*}
	\includegraphics{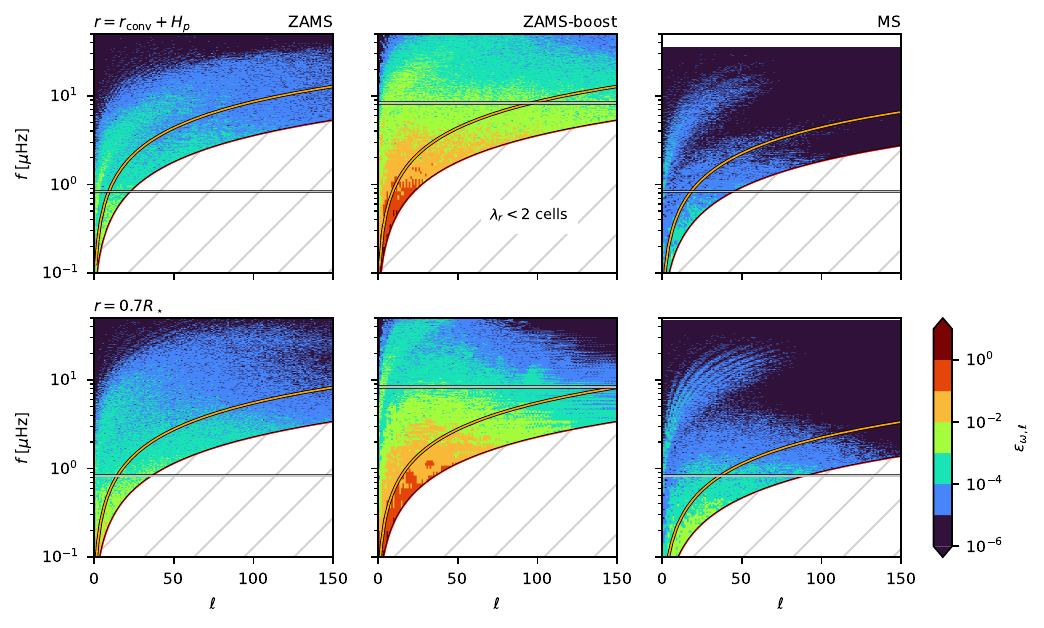}
	\caption{
		The nonlinearity parameter $\epsilon \equiv k_h u_h / \omega$ from \citet{press81}.
		The top panels are computed one pressure scale height above the convective-radiative boundary, while the bottom panels are computed at $0.7R_\star$.
		The horizontal line indicates the convective frequency of the model.
		The orange and red lines indicate waves with a radial wavelength of five and two cells in the simulation respectively.
		Waves cannot be resolved with fewer than two cells per wavelength, so this region has been blanked and diagonally hatched.
	}
	\label{fig:non-lin}
\end{figure*}

\begin{figure*}
	\begin{subfigure}{\linewidth}
		\includegraphics{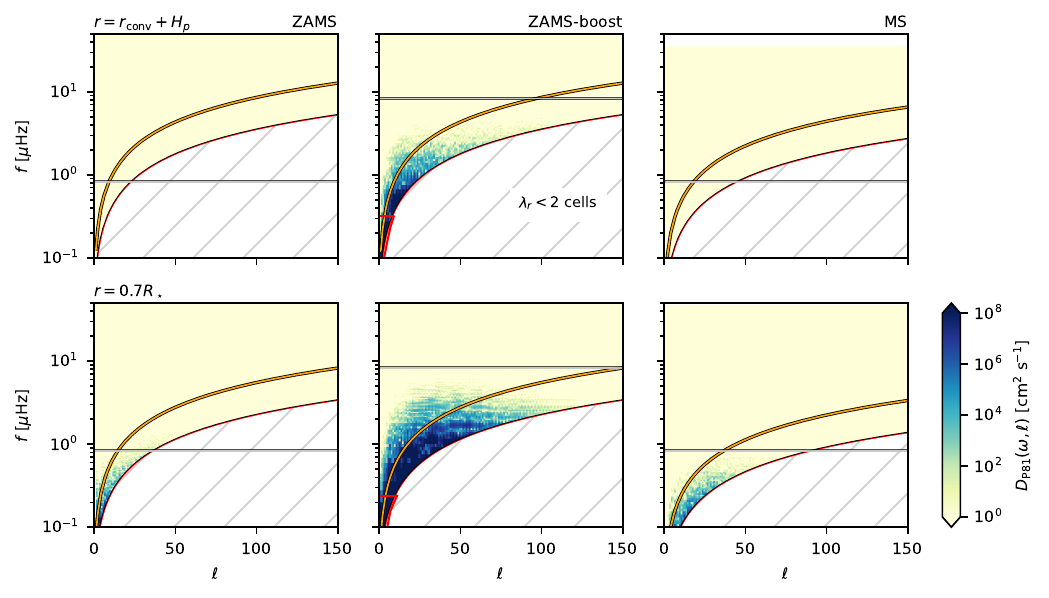}
		\caption{
			Diffusion coefficients predicted by \citet{press81}, computed using the wave amplitudes from the MUSIC simulations as \eqref{eq:diffusion-p81}.
		}
		\label{fig:diffusion-P81}
	\end{subfigure}
	\begin{subfigure}{\linewidth}
		\includegraphics{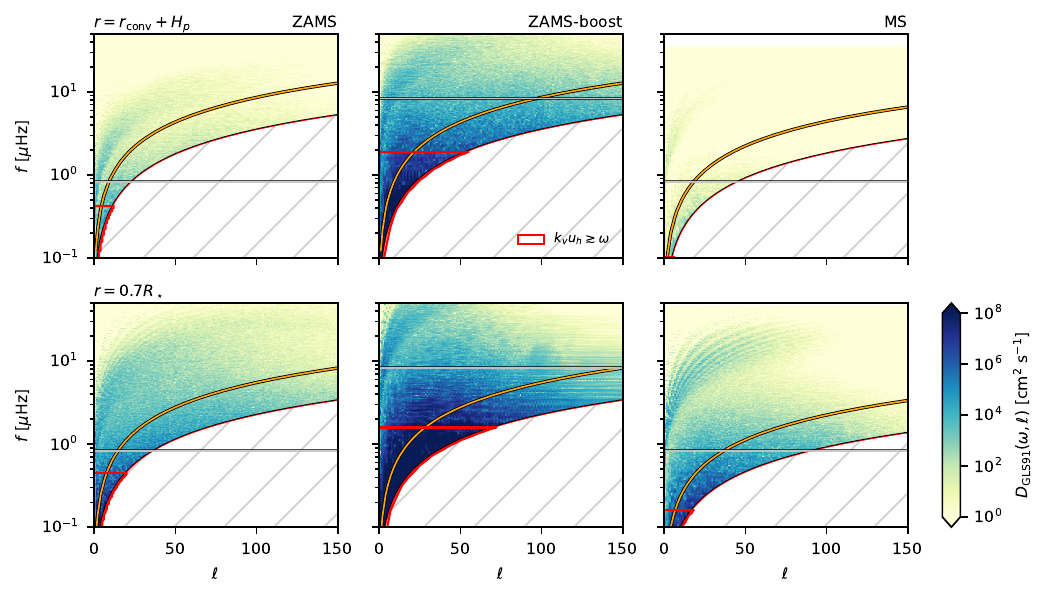}
		\caption{
			Diffusion coefficients predicted by \citet{lopez91}, computed using wave fluxes (\eqref{eq:fwave}) from the MUSIC simulations as \eqref{eq:diffusion-gls91}.
			The regions outlined in red indicate where the criterion for mixing (\eqref{eq:criterion-gls91}) is approximately satisfied.
		}
		\label{fig:diffusion-GLS91}
	\end{subfigure}
	\caption{
		Diffusion coefficient associated with each \gls*{igw} mode obtained from the power spectrum (see \secref{sec:waves}), according to two analytical models of wave mixing.
		The top row of each figure is computed one pressure scale height above the convective-radiative boundary, while the bottom row is computed at $0.7R_\star$.
		The horizontal lines show the convective frequency of each simulation, while the hatched regions indicate modes which cannot be properly resolved in our simulations, as in \figref{fig:non-lin}.
	}
	\label{fig:diffusion-theory}
\end{figure*}

The wave analysis described in \secref{sec:waves} provides all quantities needed to calculate the nonlinearity parameter $\epsilon$ and the diffusion coefficients based on the two mechanisms described in \secref{sec:mixing-models}, for the waves excited in the simulations (\tabref{tab:simulations}).
Note that, since the theoretical predictions mentioned apply to monochromatic waves, we can only evaluate them for each pair of $(\ell, \omega)$ from the spectra of waves --- deriving a total diffusion coefficient profile along the radius would require an understanding of how the contribution to diffusion from each wave of the spectrum would combine.
In fact, \citet{montalban94} takes the wave damping due to thermal diffusion derived in \citet{press81} and develops it to include effects of the excitation of `ripples' (i.e.\ a spectrum of \glspl*{igw} concentrated in space and propagating along radius) by turbulent fluid elements arriving at the convective boundary, an idea which \citet{montalban96,montalban00} use to determine lithium abundances in stars with solar metallicity and masses in the range of $0.7M_\odot$ to $1.2M_\odot$.
We do not employ their work here due to the additional complexity, but we can nonetheless comment on the typical magnitudes of diffusion coefficients obtained from \eqref{eq:diffusion-p81} and \eqref{eq:diffusion-gls91}.
\figref{fig:non-lin} shows the amplitude of the nonlinearity parameter in $\ell - \omega$ diagrams.
The parameter is shown for three different simulations in \tabref{tab:simulations}, each at two selected radii --- one a pressure scale height above the convective-radiative boundary, and another towards the outer boundary of the simulated domain.
The wave frequencies and angular degrees which we are able to resolve are limited by the time interval $\Delta t_{\rm rec}$ between outputs of the velocity field, and the grid resolution $\Delta r$ of the simulations.
Wave frequencies above the Nyquist frequency $f_s = 1/(2\Delta t_{\rm rec})$ of the simulations' outputs also cannot be resolved.
Representing wave motion requires at least two grid cells per wavelength so, to avoid misinterpretation, regions where $\lambda_r \lesssim 2\Delta r$ are shown as diagonal hatching.
\figref{fig:non-lin} indicates that the majority of the waves are in the linear regime, with $\epsilon$ reaching above $\sim 10^{-2}$ only at frequencies below the convective turnover frequency of each simulation, where \figref{fig:power-spec} shows there is considerably less power.
However, such low values of $\epsilon$ implies that the \citet{press81} mixing mechanism would be weak in these simulations, and predict very small diffusion coefficients due to the steep dependence of $D_{\rm P81}$ on $\epsilon$.
Indeed, \figref{fig:diffusion-P81} shows the diffusion coefficients predicted by \citet{press81} are extremely small in our simulations, rarely exceeding $1 \si{\centi\meter\squared\per\second}$.
Even towards the outer boundaries where \glspl*{igw} are amplified by the decrease in density, the non-boosted simulations give negligible diffusion coefficients.
Significant diffusion is only predicted in the artificially boosted model ZAMS-boost towards the outer boundary, and only for frequencies below that of the convective frequency of the simulation $\omega < 2\pi f_{\rm conv}$ where we would expect few \glspl*{igw}, though we cannot rule out that some convective motions could excite waves at these low frequencies.
Diffusion coefficients as small as those in \figref{fig:diffusion-P81} cannot be detected (at least, not reliably) in our simulations, as they would correspond to spreading of material over a distance smaller than a grid cell for the simulation time considered here.
We will provide more details on this restriction in \secref{sec:tracers}.

\figref{fig:diffusion-GLS91} shows the diffusion coefficients according to \citet{lopez91}, computed as per \eqref{eq:diffusion-gls91}.
These diffusion coefficients are larger, and certainly become relevant in the boosted simulation, and in the non-boosted simulations towards the outer boundary due to increased wave amplitudes there.
Diffusion coefficients reach $\sim 10^4 \si{\centi\meter\squared\per\second}$ at $0.7R_\star$ of the non-boosted simulations, however they are still mostly negligible just above the convective core where the need for extra mixing is most relevant.
\glspl*{igw} in the ZAMS simulation contribute to a diffusion coefficient of only $D \sim 10^2 \si{\centi\meter\squared\per\second}$ at a pressure scale height above the core of the ZAMS model, though even this is likely an over-estimate, as standing waves are strengthened by the numerical reflective outer boundary so could artificially enhance the result via large amplitudes in \eqref{eq:diffusion-gls91}.
The evolved MS simulation yields smaller diffusion coefficients still, due to the wave damping that occurs in the $N$-peak layer atop the core, as shown in \secref{sec:power-spec-ms} and \citet{morison24}.
Furthermore, the mixing criterion \eqref{eq:criterion-gls91} is only satisfied for frequencies well below the convective frequency of each simulation, providing further restrictions on the waves that could contribute to mixing via this theory.
Hence, these results predict that mixing atop the core due to sub-wavelength turbulence would be very weak in both zero-age and evolved \gls*{ms} massive stars.

%% file: src/05-tracers.tex

\section{Mixing of Lagrangian tracers}\label{sec:tracers}

In the following, we embed tracer particles in the numerical simulations performed for this work, to measure mixing of material which could be linked to the predictions made in \secref{sec:wave-mixing}.
Since tracer particles follow Lagrangian trajectories, we can use the statistics of their motion over sufficient time periods to characterise the mixing in the simulations, a technique that has been used in multiple recent studies \citep[e.g.][]{rogers17,cunningham19,higl21,varghese23}.

\subsection{Lagrangian tracers}

We track Lagrangian trajectories in the flow using a collection of tracer particles,
labelled by an index $p \in \{1, \ldots, N_P\}$.
Each particle's position $\bm x_p$ is advected by the local flow velocity:
\begin{equation}
    \diff{\bm x_p}{t} = \bm v(\bm x_p(t), t).
    \label{eq:particle-advection}
\end{equation}
The velocity field from the hydrodynamical simulation is only available at discrete space points $i$ (on the grid) and time points $n$, so interpolation is required to obtain $\bm v$.

At each time point $n$, we first define a spatially-continuous velocity field $\tilde{\bm v}_n(\bm x)$,
using either linear or cubic (Hermite) spatial interpolation of the discrete velocities $\bm v_{i,n}$.
For the purposes of our conclusions, we find overall similar results with the linear and Hermite spatial interpolation schemes,
and will present results for linear interpolation unless otherwise stated.

We integrate equation \eqref{eq:particle-advection} for the particles using an explicit method. Our choice of time step for MUSIC's implicit hydro solver implies that no particle can cross more than one cell over one hydro time step.
We evolve the positions $\bm x_p$ in time using $\tilde{\bm v}_n(\bm x)$ and a second-order Runge-Kutta scheme (Heun's method):
\begin{align}
    \bm x'_{p,n+1} &\coloneqq \bm x_{p,n} + \Delta t \, \tilde{\bm v}_n(\bm x_{p,n})
    \label{eq:timestep-pred}
    \\
    \bm x_{p,n+1} &= \bm x_{p,n} + \Delta t \,
        \frac{1}{2} \left( \tilde{\bm v}_n(\bm x_{p,n}) + \tilde{\bm v}_{n+1}(\bm x'_{p,n+1}) \right)
    \label{eq:timestep-heun}
\end{align}
where $\Delta t = t_{n+1} - t_n$ is the time interval between the two consecutive time points.
We have also implemented the simple first-order Euler method,
which corresponds simply to using the predictor step of \eqref{eq:timestep-pred} above
as the final estimate of the new particle position.
We present results using Heun's method, unless otherwise stated.

MUSIC can integrate particle tracers as a post-processing step;
in this case, $\Delta t$ is set by the interval between successive output snapshots $\Delta t_{\rm rec}$.
MUSIC can also integrate particles online, i.e. at the same time as the numerical time stepping;
online integration allows reaching a much smaller $\Delta t$, and is therefore more accurate.
Further details are given in Appendix~\ref{appendix:online-particles}.
In results shown, we integrate the trajectories in post-processing with the same $\Delta t_{\rm rec}$ as that used to compute the power spectra in \figref{fig:power-spec}.

\subsection{Characterization of radial particle transport}\label{sec:particle-method}

Authors have proposed different methods to characterize vertical Lagrangian transport in stellar simulations.
\cite{freytag96,cunningham19} followed particles that are initially seeded along horizontal shells.
This method is straightforward, and allows for accurate tracking of vertical motions of the particles,
since their starting radii are known exactly.
However, properties of the particle transport can then only be assigned to a particular radius through the initial seed location of the particles.
This prevents benefiting from ergodicity, as one cannot average over trajectories starting at a same radius but at different times in the flow.

The method proposed by \cite{rogers17}, which we also adopted in this work, overcomes this limitation.
The idea is to look at trajectories over time intervals $[t, t+\tau]$,
and assign trajectories to a radius $r$ by interpolating from their starting radii.

Taking the radial coordinate $R_p(t) \coloneqq r(\bm x_p(t))$ of each particle $p$ as a function of time,
we consider the radial displacement $\delta R_p(t, \tau)$ of particle $p$
over the course of the trajectory:
\begin{equation}\label{eq:tracer-displacement}
    \delta R_p(t, \tau) \coloneqq R_p(t+\tau) - R_p(t).
\end{equation}
If the transport is purely diffusive, it can be characterized by the statistics
of the squared displacement (second moment) $\delta R_p^2$.

At a given time $t$, each particle's radial position $R_p(t)$ will be unique, so to obtain statistics over many particles, we interpolate any derived quantities from the starting radius of each particle's trajectory to arbitrary grid points $r$.
This is done by weighting each trajectory's contribution to a given statistic by its initial radial distance from the grid point $r$, using a cubic spline kernel $w(r - R_p(t))$.
Accordingly, like \citet{rogers17}, we define the number of trajectories, displacement, and squared displacement at radius $r$ as:
\begin{align}
    n(r, \tau) &\coloneqq \left< \sum_p w(r - R_p(t)) \times 1 \right>_t, \\
    P(r, \tau) &\coloneqq \left< \sum_p w(r - R_p(t)) \times \delta R_p(t, \tau) \right>_t, \\
    Q(r, \tau) &\coloneqq \left< \sum_p w(r - R_p(t)) \times \delta R_p(t, \tau)^2 \right>_t,
\end{align}
where the time averages integrate over all starting points $t \in [0, T-\tau]$
for sub-trajectories of duration $\tau$.

The advantage of the time average is to reduce the variance in the estimation of moments,
assuming that the transport properties are ergodic and that the time span $T$ is much longer than the autocorrelation timescales of the flow.
For \glspl*{igw}, this requires at least $T \gg \tau_\text{conv}$.
Note that for small values of $T$, or with non steady-state flow, the time average can hide a complex time dependence and produce misleading results.

Coming back to the study of diffusion processes, the moments of $\delta R_p$ can now be computed as:
\begin{align}\label{eq:rm2017}
    \left<\delta R_p\right> &= \frac{P}{n},  & \left<\delta R_p^2\right> &= \frac{Q}{n}, &
    \sigma^2(r,\tau) &\coloneqq \operatorname{var} \delta R_p = \frac{Q}{n} - \frac{P^2}{n^2},
\end{align}
where $\sigma^2(r,\tau)$ is the \emph{mean squared displacement}, and follows the spread of particles initially at $r$ while compensating for any average vertical drift of the particles by subtracting the $P$ term.

As argued by \cite{rogers17}, for particles in an oscillatory flow with vertical diffusive transport (Brownian motion) with diffusion coefficient $D$, we expect
\begin{equation}
	\label{eq:msd}
    \sigma^2(\tau) \approx C + 2D\tau, \quad \tau \gg \text{flow oscillation period}.
\end{equation}
For an initial duration $\tau$, the particles will be displaced in a `ballistic' manner by the oncoming wavefronts, as with a boat hitting the front of another's wake.
For an ensemble of waves with a spectrum of frequencies and uniformly distributed phases, the ballistic motion of particles will occur over a duration of approximately the longest wave period, until all the wave peaks have passed and the particles oscillate within a region set by the mean wave amplitude.
Hence, for the long durations in \eqref{eq:msd}, the constant $C$ corresponds to the initial squared displacement produced by the ballistic regime, set by the mean squared wave amplitude and thereby related to the RMS velocity of the flow.
The `oscillation' cartoon in \figref{fig:msd-profile} illustrates this behaviour.
To measure diffusion, the ballistic regime should be ignored by subtracting the constant $C$ from $\sigma^2(\tau)$ before fitting a linear relationship, and the remaining oscillatory motion will average to zero, given a long enough duration.

\subsection{Applying the particle method to our simulations}\label{sec:particle-results}

\begin{figure}
	\begin{subfigure}{\linewidth}
		\includegraphics[width=\linewidth]{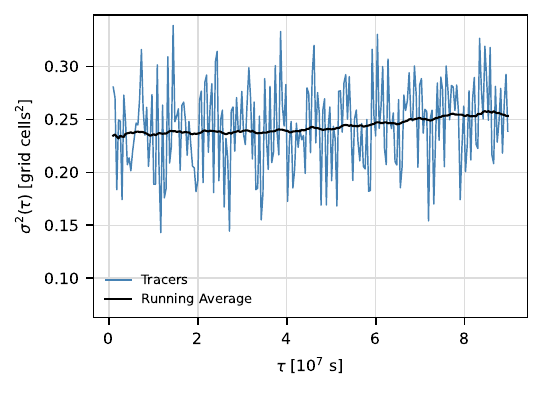}
	\end{subfigure}
	\hfill
	\begin{subfigure}{\linewidth}
		\includegraphics[width=\linewidth]{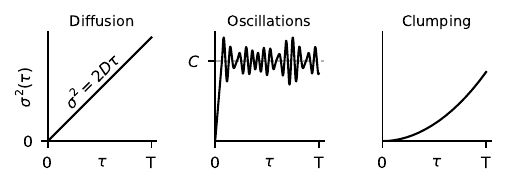}
	\end{subfigure}
	\caption{
		\emph{Top panel}: Mean squared displacement as a function of lag time for the non-boosted ZAMS simulation at $r=0.6R_\star$.
		\emph{Bottom panel}: Cartoons of mean squared displacement profiles that would arise due to three different processes acting on the tracers --- pure diffusion, an ensemble of oscillations with a spectrum of frequencies, and a model of the particle clumping artefact we observe in the simulations (\secref{sec:particle-method}, Appendix \ref{appendix:msd}).
	}
	\label{fig:msd-profile}
\end{figure}

\figref{fig:msd-profile} shows the mean squared displacement computed for the non-boosted ZAMS simulation at $0.6R_\star$ using the method of \citet{rogers17} with $10^6$ particles distributed across the radiative zone.
We use a similar number of particles for all the simulations performed in this work.
It is clear that the oscillatory motion of the particles dominates, made apparent by the large mean squared displacements at very small lag $\tau$, on the order of the wave periods in our simulations.
Superimposed on this signature of wave motion, we do observe a positive correlation between $\sigma^2$ and $\tau$.
However, it appears that the relationship is not linear, and hence not characteristic of regular diffusion.
We observe this behaviour in most of the simulations performed in this work.

\begin{figure*}
	\begin{subfigure}{\linewidth}
		\includegraphics[width=\linewidth]{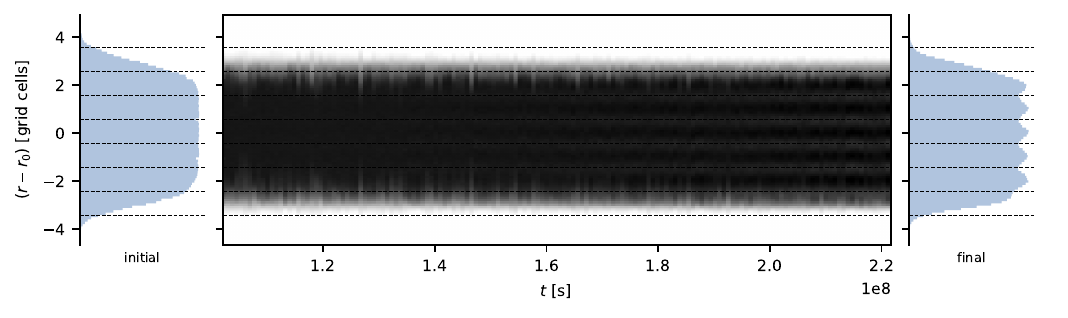}
	\end{subfigure}
	\begin{subfigure}{\linewidth}
		\includegraphics[width=\linewidth]{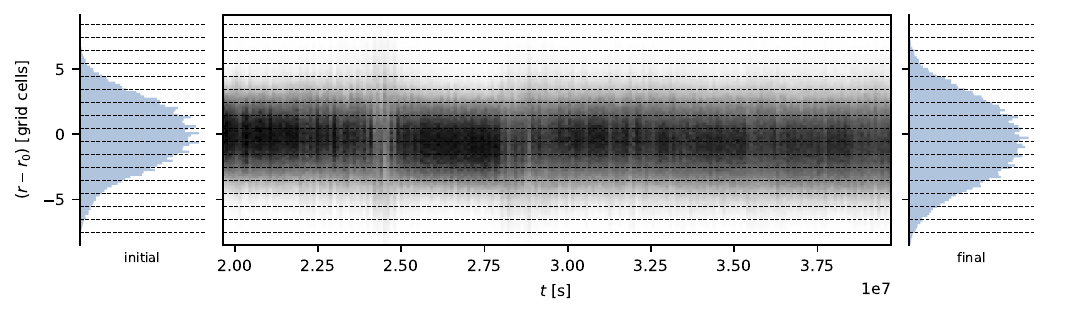}
	\end{subfigure}
	\caption{
		Histograms computed from tracer particles initialised at time $t=t_{\rm steady}$ (from \tabref{tab:simulations}) as top hat distributions centred around $r=0.6R_\star$ for simulation ZAMS (\emph{top}) and $r=0.7R_\star$ for simulation ZAMS-boost (\emph{bottom}).
		The tracers are binned by their radial positions at a given time, and the histograms are stacked together along a time axis.
		The MUSIC radial grid cell edges are shown as horizontal dashed lines.
	}
	\label{fig:particle-histogram}
\end{figure*}

A likely explanation for this is an artefact we observe in the motion of the particles.
Over $\sim 100$ periods of wave motion, the particles slowly collect towards the centres of the simulation cells, as shown by \figref{fig:particle-histogram} for the ZAMS simulation at $r=0.6R_\star$.

This ``clumping'' artefact shows that the particle density does not exactly follow the mass of the hydrodynamical Eulerian fields over long times.
We suggest that its root cause is that the spatial interpolation of velocity fields to particle locations
does not exactly preserve the divergence of the velocity field.
This is a subtle but well-known problem in particle tracking, for which elaborate solutions have been proposed
\citep[see e.g.][]{zhangGeneralMassConsistency2004,mcdermottParabolicEdgeReconstruction2008,popovAccurateTimeAdvancement2008}, though not developed for stellar interiors.
Different techniques based on probabilistic tracers have also been formulated \citep{genelFollowingFlowTracer2013,cadiouAccurateTracerParticles2019}.
In our simulations, this problem is present with both linear and Hermite spatial interpolations,
and is most pronounced for simulations that exhibit a particle displacement amplitude of less than or similar to a MUSIC grid cell.

In addition, we found that the choice of time integrator and particle tracking time step can aggravate the issue.
Many explicit ODE methods (such as Euler and Runge-Kutta methods) typically become unstable for oscillatory solutions
if the time step is too large compared to the oscillation period,
because they are not A-stable \citep[see e.g.][]{butcherNumericalMethodsOrdinary2016}.
We note that particles will feel the effects of a whole spectrum of waves, including high-frequency \glspl*{igw}.
This can cause artificial dispersion of the particles,
for example when using offline particle with velocity field snapshots that are not frequent enough,
or if the integration period is long enough for small errors to accumulate.
First-order methods such as explicit Euler are particularly susceptible,
and we have indeed noted extra apparent particle dispersion with the Euler method
when experimenting with snapshot frequency.

\begin{figure}
	\includegraphics[width=\linewidth]{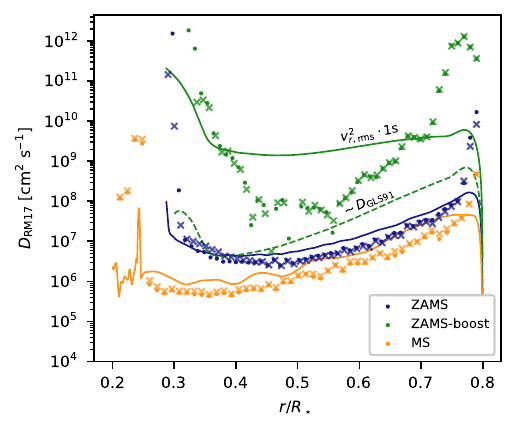}
	\caption{
		$D$ from \eqref{eq:msd}, following the tracer particle method of \citet{rogers17} (described in \secref{sec:particle-method}) for three different simulations performed in this work.
		Results using linear spatial interpolation of the velocities are shown as dots, and those using Hermite interpolation are shown as crosses.
		The solid lines show $v_{r, \rm rms}^2 \cdot 1\si{\second}$ for each simulation.
		The dashed line shows \eqref{eq:diffusion-gls91} evaluated for ZAMS-boost at $\ell = 1, \omega = 2\pi f_{\rm conv}$, with the wave flux estimated with the RMS radial velocity as in \eqref{eq:fwave-vrms}.
		A similar prediction from \eqref{eq:diffusion-p81} is not shown here, since the resulting diffusion coefficients are smaller than the lower limit on the $y$-axis.
	}
	\label{fig:diffusion-rm17}
\end{figure}

We can identify the effect of the clumping artefact in the mean squared displacement profile $\sigma^2(\tau)$.
Using a simple model, we can show that $\sigma^2(\tau)$ due to this artefact is not linear in $\tau$, and instead shows a gradient increasing with $\tau$ (see Appendix \ref{appendix:msd}), which helps to distinguish it from diffusive behaviour.
In cases where the clumping artefact is significant, the mean squared displacements indeed resemble this result, as in \figref{fig:msd-profile}.
Continuing to follow the method outlined in \secref{sec:particle-method}, we fit \eqref{eq:msd} to the mean squared displacement profiles to obtain $D$.
For the models, time duration $T$, and range of durations $\tau$ used in this work, $D$ coincidentally appears closely related to the RMS radial velocity of the simulation as $D \sim v_{r, \rm rms}^2$ (\figref{fig:diffusion-rm17}), despite being entirely dependent on the strength of the numerical artefact.
We stress that this result should \emph{not} be interpreted physically.

\begin{figure}
	\includegraphics[width=\linewidth]{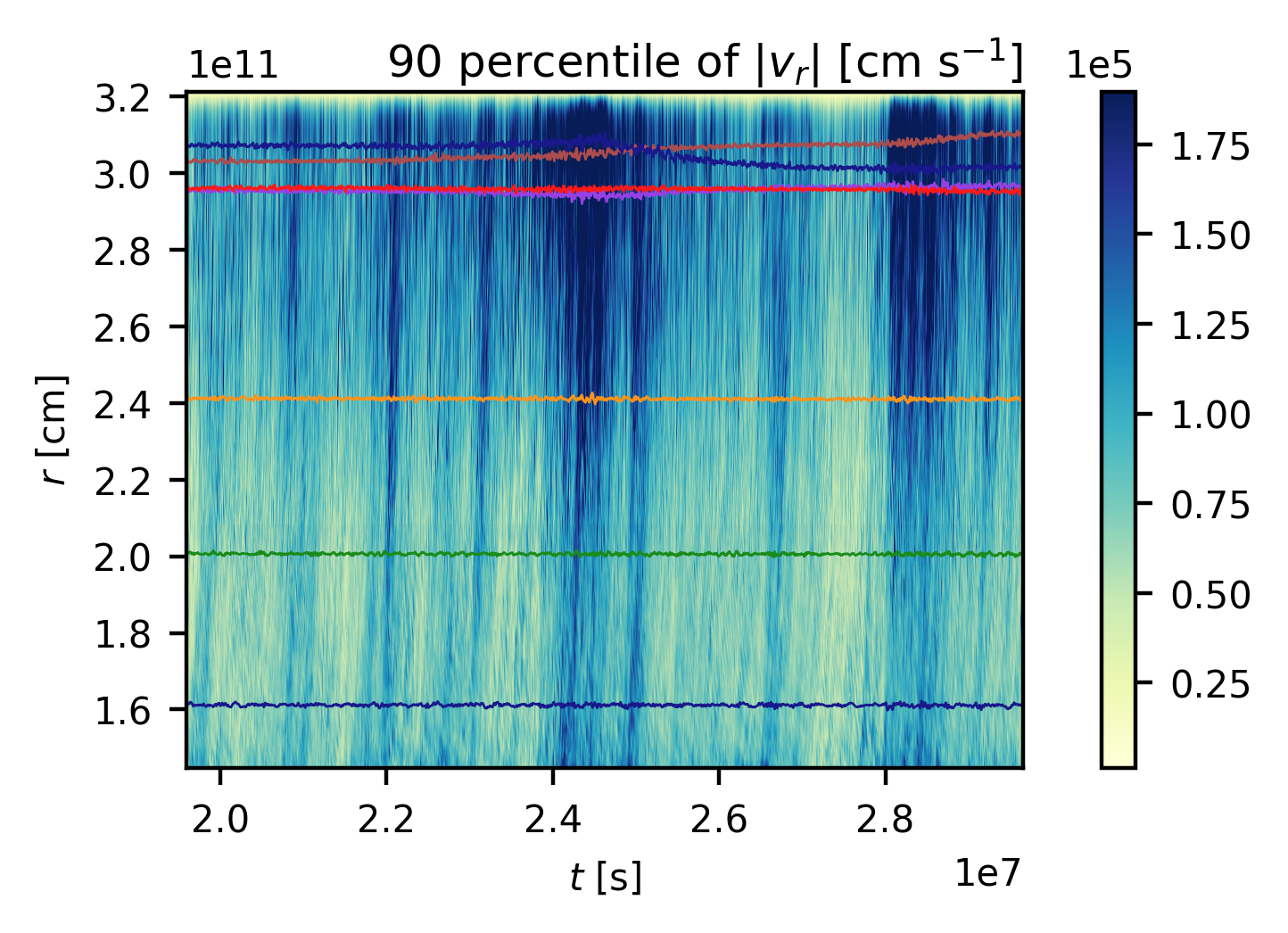}
	\caption{
		The 90th percentile (along the theta direction) of the absolute radial velocity of ZAMS-boost, as a function of time and radius.
		Intermittent groups of strong \glspl*{igw} are visible, propagating from the convective boundary outwards.
		The lines show trajectories of single tracer particles selected at various depths throughout the radiative zone.
	}
	\label{fig:vr-quantile}
\end{figure}

The ZAMS-boost simulation gives a different result, since it is the only simulation which does not exhibit the particle clumping.
We speculate that this is due to the vertical wave displacements $\xi$ being much larger than a grid cell, thereby weakening the influence of the clumping artefact, as can be seen in the particle histogram in \figref{fig:particle-histogram}.
The diffusion coefficients in the majority of the radiative zone are smaller than $v_{r, \rm rms}^2 \cdot 1\si{\second}$ by up to two orders of magnitude, differing from the non-boosted simulations and suggesting that the mixing due to \glspl*{igw} in our simulations is weaker than that in \citet{rogers17}.
However, instead of a gradual spreading out, these particles intermittently undergo large displacements (with respect to typical wave amplitudes) over short periods of time, indicating at the presence of events isolated in time.
The impact of such events can be seen in \figref{fig:particle-histogram} for the ZAMS-boost simulation at $r=0.7R_\star$, with sudden changes in the distribution of particles at times $t \sim 2.4\times 10^7\si{\second}$ and $t \sim 2.8 \times 10^7 \si{\second}$.
Indeed, the boosted simulation shows collections of high amplitude \glspl*{igw} appearing at these times and originating from the top of the convective core (\figref{fig:vr-quantile}), possibly due to the extreme events of penetrating convection discussed in \citet{baraffe23}.
A few selected particles' trajectories are shown in \figref{fig:vr-quantile}, which highlights their movement during these intermittent groups of \glspl*{igw}.
Whether or not the temporally and/or spatially localised nature of the waves is important, or whether the large amplitudes are the sole reason for the displacement of the particles is so far unclear.
In this work, we will not speculate further on the effect of these groups of waves on particle mixing.
We do note that the motion of particles in the outer $0.05R_\star$ is possibly influenced by boundary effects and strong wave reflections.

%% file: src/06-discussion.tex

\section{Discussions and conclusions}\label{sec:discussion}

We have performed simulations of $20M_\odot$ stars, with both \gls*{zams} and evolved mid-\gls*{ms} models, as well as a \gls*{zams} simulation with an artificially boosted luminosity and radiative diffusivity.
This work firstly aimed to evaluate diffusion coefficients based on the predictions of \citet{press81} and \citet{lopez91} for chemical mixing due to \glspl*{igw}, using the wave spectra and fluxes from our numerical simulations.
Here we comment on the typical magnitudes of these predictions, to determine if and where the mechanisms might be most important in the stars.
\eqref{eq:diffusion-p81}, using theory from \citet{press81}, predicts little diffusion ($D < 1 \si{\centi\metre\squared\per\second}$) in our simulations (\figref{fig:diffusion-P81}), especially near the bottom of the radiative zone where mixing would have the greatest effect on stellar evolution, transporting material across the convective boundary.
Artificially boosting the luminosity and radiative diffusivity of the simulation by a factor of $10^3$ does raise the predicted diffusion coefficients for \glspl*{igw} at low frequencies, but since the convective frequency is also raised, the theory still predicts negligible mixing for waves with $\omega \gtrsim 2\pi f_{\rm conv}$.
With no significant diffusion for waves of frequencies near or above the convective frequency of the simulation, we cannot conclude that the \citet{press81} mixing mechanism should have any significant effect in the models we consider in this work, nor that they could be measured in our numerical simulations.

The magnitude of the diffusion coefficient is predicted to be larger when using the suggestion of \citet{lopez91} (\eqref{eq:diffusion-gls91}).
The \gls*{zams} model shows diffusion coefficients of the order $10^2 \si{\centi\metre\squared\per\second}$ at a pressure scale height above the convective core (\figref{fig:diffusion-GLS91}), increasing to $10^4 \si{\centi\metre\squared\per\second}$ at $r=0.7R_\star$ due to the fall in density leading to increased wave amplitudes.
However, the evolved \gls*{ms} model gives rise to smaller diffusion coefficients by around two orders of magnitude due to the damping of \glspl*{igw} in the peak of Brunt-V\"ais\"al\"a frequency at the base of the radiative zone, as found by \citet{morison24}.
Even these coefficients are likely overestimated compared to real stars, as there are large contributions to the wave power from strong standing g-modes in the simulations (\figref{fig:power-spec}), which are certainly enhanced in our simulations by the hard reflective outer numerical boundary (see the later discussion of damping layers).
The most significant diffusion coefficients in \figref{fig:diffusion-GLS91} typically occur at frequencies of the order of the convective frequency or smaller, which correspond to \glspl*{igw} with small radial wavelengths (as implied by the dispersion relation, \eqref{eq:igw-dispersion}).
Additionally, the criterion in \eqref{eq:criterion-gls91} is only satisfied for low frequencies well below the convective frequency of each simulation (\figref{fig:diffusion-GLS91}), where the shear induced by the wave would be sustained for a sufficient period of time.
While this criterion is derived by \citet{lopez91} using only approximate arguments, it hints that even strong waves above the convective frequency may not contribute to mixing via this mechanism.
Resolving the flows on spatial scales corresponding to the short wavelengths of these \glspl*{igw} is difficult for hydrodynamic stellar simulations.
\figref{fig:diffusion-GLS91} indicates that the simulations performed here have approximately 5 cells per radial wavelength at $0.5 \si{\micro\hertz}$, which is not sufficient to resolve a Kelvin-Helmholtz instability.
Hence, we do not expect to observe the mixing mechanism of \citet{lopez91} in this work.
\\

\begin{table}
	\caption{
		Minimum detectable diffusion coefficients if we require to see a standard deviation $\sigma$ of one MUSIC grid cell in the distribution of material initialised at a single point in space over the simulation times listed in \tabref{tab:simulations}.
	}
	\label{tab:Dmin}
	\centering
	\begin{tabular}{lc}
		\hline \hline
		Model & $D_{\rm min}$ (\si{\centi\meter\squared\per\second}) \\
		\hline
		ZAMS & $1.7 \times 10^8$ \\
		ZAMS-boost & $1.4 \times 10^9$ \\
		MS & $1.6 \times 10^8$ \\
		\hline
	\end{tabular}
\end{table}

In order to measure mixing occurring in our simulations, we employed the Lagrangian tracer particle methods of \citet{rogers17}.
By embedding tracers into the flow, we can use the statistics of their displacements over the course of the simulation to quantify the transport, assuming a diffusive process.
We wish to stress that tracer particles are only able to trace diffusive effects that are mediated by the hydrodynamical velocity field (i.e., mixing through ``stirring''); in particular, passive tracer particles advected by \eqref{eq:particle-advection} cannot feel microscopic diffusion.
Although, since the mechanisms of \citet{press81} and \citet{lopez91} involve the displacement of fluid parcels, we would expect to be able to probe them via the hydrodynamical velocity field.

However, we argue here that tracer particles have difficulties acting as a reliable measure for mixing by \glspl*{igw} in hydrodynamic stellar simulations.
After $\sim 50$ convective timescales, we begin to see the tracer particles collecting in the cell centres of the grid used for the hydrodynamical simulation.
While sophisticated interpolation and integration methods \citep[e.g.][]{popovAccurateTimeAdvancement2008,wangAdvantagesConservativeVelocity2015} may lessen the issue, such methods have not been utilised in any prominent studies of mixing within stars.
Supporting this conclusion, the `boosted' ZAMS simulation does not suffer from the same artefact in the time span analysed, since the larger wave amplitudes lead to particle displacements spanning multiple cells, thereby weakening the influence of the sub-grid velocity interpolation.
In addition to artefacts introduced by interpolation, time integration over many periods in oscillatory flows can accumulate errors and contribute to the clumping effect or artificial dispersion of the particles.
The use of a significantly reduced timestep for integration of the particle trajectories and alternative methods of velocity interpolation did not remove the effect (see Appendix~\ref{appendix:online-particles}), and cubic spline interpolation can actually aggravate it.
Thus, we stress the crucial role of both the spatial interpolation and time integration schemes in obtaining reliable estimates for the diffusion coefficients in oscillatory flows.

These issues highlight important limitations of particle methods in hydrodynamical stellar simulations.
Diffusion coefficients predicted by theory in \secref{sec:wave-mixing} are small, so numerical simulations need to be run for a long time without any artefacts appearing over small spatial scales.
Previous studies may not have suffered from this numerical artefact, but measured diffusion coefficients of $D \sim v_{\rm rms}^2$ imply rapid mixing in the simulations which is incompatible with the theoretical predictions of \secref{sec:wave-mixing}.
\tabref{tab:Dmin} shows the minimum diffusion coefficients required to create a spreading of material comparable to a single grid cell of our simulations.
Regardless of how accurate the particle integration is, measurements of diffusion weaker than this are not robust as they are based on interpolated sub-grid-cell effects, given the resolution and time duration used in this work.
As such, caution is required when attempting to quantify mixing using tracer particles in hydrodynamic simulations, especially since the resulting diffusion coefficients in \figref{fig:diffusion-rm17} appear reasonable without deeper examination of the particle trajectories.
Results derived from such methods are already being used as realistic mixing profiles with one-dimensional stellar evolution codes \citep[e.g.][]{pedersen2018,michielsen2023,li2023}, so it is clear that there is a need for more robust results and connections to theory.\\

As discussed, the tracer particles in the artificially boosted simulation (ZAMS-boost) do not exhibit any obvious numerical artefacts.
However, the recovered diffusion coefficients from the tracers bear no relation to any of the discussed mechanisms, nor to the prediction of $D \sim v_{r, \rm rms}^2 \cdot 1\si{\second}$ from \citet{rogers17,varghese23} (\figref{fig:diffusion-rm17}).
For this boosted simulation, we recover diffusion coefficients smaller than $v_{r, \rm rms}^2 \cdot 1\si{\second}$ (except from the outer $0.1R_\star$) by up to two orders of magnitude.
We do note the difference in viscosity between this study and that of \citet{rogers17, varghese23}.
The hydrodynamic code utilised in those studies employs an explicit uniform viscosity across the entire domain.
On the other hand, MUSIC uses the \gls*{iles} approach, where dissipation is introduced locally in the flow based on grid resolution and flow morphology.
With this approach, momentum is often dissipated slower than with a stabilizing explicit viscosity.
Indeed, the non-boosted simulations pictured in \figref{fig:snapshots} have visibly weaker effective dissipation than the $20M_\odot$ simulation of \citet{varghese23}, with dissipation acting to damp waves of shorter wavelength.
Resolution convergence tests could assess the impact of numerical dissipation, but are beyond the scope of this study.
Such tests would also be supported by work currently in progress to assess dissipation in the ILES scheme of MUSIC.
The effects of viscosity on mixing by \glspl*{igw} are not discussed in either \citet{press81} or \citet{lopez91}, though viscosity tends to strongly damp waves of low frequency \citep{alvan13} which would only reduce the diffusion predicted by \eqref{eq:diffusion-p81} and \eqref{eq:diffusion-gls91}.
Additionally, we would argue that one cannot reliably extrapolate these diffusion coefficients to stellar evolutionary timescales due to the infrequent appearances of high amplitude waves over the simulated period of time.
Such behaviour is not compatible with a diffusion model over timescales covered by the hydrodynamic simulations, since the occasional appearance of groups of strong waves breaks the time-independence of the particle motions.

The diffusion predicted in ZAMS-boost by \eqref{eq:diffusion-p81} (shown in \figref{fig:diffusion-P81}) is far too weak to be detected reliably over the timespan of our simulations, despite the particles not suffering from the same clumping artefact as in the non-boosted simulations.
While the diffusion coefficients predicted using \eqref{eq:diffusion-gls91} (shown in \figref{fig:diffusion-GLS91}) are significant for ZAMS-boost, they exhibit no similarity to those measured using tracer particles (\figref{fig:diffusion-rm17}).
As a result, the mechanism causing the ``mixing''\footnote{The tracer particle method used in this work can not discern true diffusion-like mixing of tracers from any other motion aside from collective drift, so we use the word ``mixing'' lightly here to mean any motion of the tracers that contributes to the measured diffusion coefficient.} of the tracer particles cannot be matched with theory.

We have also performed simulations with a damping layer at the outer numerical boundary, as listed in \tabref{tab:damped-simulations}.
Since g-modes are present in real stars, the simulations without a damping layer serve as a good experiment to determine mixing profiles for 1D stellar evolution codes, but the theories for \gls*{igw} mixing introduced in \secref{sec:mixing-models} apply only to propagating waves.
Using a damping layer thus provides more relevant results for comparison with these theories.
We find that the standing modes in these simulations are greatly reduced in power, compared with those in the undamped simulations.
The diffusion coefficients measured using the tracer particle method discussed in \secref{sec:tracers} again follow the trend of $v_{r, \rm rms}^2$, for the same reasons as in the non-boosted ZAMS and MS simulations without a damping layer.
Otherwise, few results are shown from these simulations since they are qualitatively similar to those of their undamped counterparts, leading to unchanged conclusions for this work.
Due to the damping, RMS velocities are lower for the damped simulations (see \figref{fig:vrms-profiles}), making it more difficult to identify any mixing processes occurring, without performing the simulation for a greater period of time and/or with higher spatial resolution.
\\

This work highlights the complexity of radial mixing by \glspl*{igw} on both the theoretical and numerical sides.
So far, there has been no conclusive demonstration that the mechanisms discussed in \secref{sec:wave-mixing} (and applied to our simulations in \figref{fig:diffusion-theory}) work in stars, and we have predicted that the mixing which would result from them would be too weak to measurably increase the extent of convective cores.
Finally, we highlight issues arising from Lagrangian tracer particle methods in measuring radial mixing due to \glspl*{igw} over long time periods in stellar simulations.
We suggest that more robust methods should be employed, to provide confidence in stellar evolution models.

%% file: src/appendix.tex

\section{Mean squared displacements due to the clumping artefact}\label{appendix:msd}
The clumping artefact observed in \secref{sec:tracers} can be modelled as tracer particles $p$ with velocities directed towards cell centres as
\begin{equation}
v_{p, \rm clump}(t) \equiv \frac{1}{T_{\rm clump}} \cdot \left[r_0 - R_p(t)\right],
\end{equation}
where $r_0$ is the nearest radial cell centre to the particle's radial position $R_p(t)$, and $T_{\rm clump}$ is the time taken for particles to clump into $e^{-1}$ of a cell width $\Delta r$.
Integrating this, we find the particle trajectories of
\begin{equation}
R_p(t) = R_{p, 0}\, e^{-t/T_{\rm clump}},
\end{equation}
where $R_{p, 0} = R_p(t = 0)$, {\it i.e.} the tracers follow an exponential decay towards the cell centres.
Hence, the radial displacement of a particle $p$ over a time $\tau$ (as per \eqref{eq:tracer-displacement}) is
\begin{equation}
\delta R_p(t, \tau) = R_{p, 0} \left(e^{-\frac{t + \tau}{T_{\rm clump}}} - e^{-\frac{t}{T_{\rm clump}}} \right)
\end{equation}
We can derive the resulting mean squared displacement profile $\sigma^2(\tau)$ by averaging over all start times $t$ within a duration $T$, and over all initial particle positions $R_{p, 0}$ (as per \eqref{eq:rm2017}, with $\langle R_{p, 0} \rangle = 0$)
\begin{align}
\sigma^2(\tau) &= \left\langle \delta R^2_{p, 0}(t, \tau) \right\rangle_{p, t} \\
&= \frac{1}{T - \tau} \int_0^{T - \tau} \frac{1}{\Delta r} \int_{-{\Delta r} / 2}^{{\Delta r} / 2} \delta R_p^2(t, \tau)\; dR_{p, 0}\; dt
\end{align}
Since the clumping artefact appears over the period $T$ of the ZAMS simulation in \figref{fig:particle-histogram}, we can take $T_{\rm clump} \sim T$ to find
\begin{equation}
\sigma^2(\beta) = \frac{(\Delta r)^2}{24} \frac{1}{1 - \beta} \left(e^{-\beta} - 1\right)^2 \left(1 - e^{2(\beta - 1)}\right),
\end{equation}
where $\beta \equiv \tau / T$.
This result is plotted in \figref{fig:msd-profile}.

\section{Impact of Boundary conditions and particle integration}\label{appendix:online-particles}

The results presented in the main body of this work are based on MUSIC simulations with reflective boundaries in the angular direction, and particle integration computed in post-processing with linear interpolation of the simulation velocity grid.
In this appendix, we assess whether the particle `clumping' artefact discussed in \secref{sec:particle-results} is reduced or eliminated by \emph{a}.\ more accurate time integration; \emph{b}.\ a higher order interpolation scheme; or \emph{c}.\ periodic boundary conditions in the angular direction.
For improved time integration, MUSIC can integrate particles `online', such that the particles are advected with each timestep of the hydrodynamical scheme.
This results in a small $\Delta t$ for \eqref{eq:timestep-heun} and thus significantly reduced time integration errors, compared with post-processing where the particle timestep is limited to the frequency at which MUSIC outputs velocity fields to disk storage.
Using a timestep as small as this, we would not expect to see significant further improvements by utilising higher-order numerical time integration methods.
For the interpolation scheme, we have additionally implemented Hermite cubic splines.
Linear interpolation (along one dimension) gives a velocity field of
\begin{equation}\label{eq:interp-linear}
\tilde{\bm v}_n(x) = \bm v_i + (\bm v_{i+1} - \bm v_i) \frac{x - x_i}{\Delta x}, 
\end{equation}
where $\Delta x$ is the grid spacing, and $x_i$ and $x_{i+1}$ are the grid points immediately to the left and right of $x$.
Hermite interpolation instead gives
\begin{equation}\label{eq:interp-hermite}
\tilde{\bm v}_n(x) = h_{00}(\gamma) \bm v_i + h_{10}(\gamma) \bm v_i' \Delta x + h_{01}(\gamma) \bm v_{i + 1} + h_{11}(\gamma) \bm v_{i+1}' \Delta x,
\end{equation}
where $\bm v_i' = (\bm v_{i+1} - \bm v_{i-1}) / (x_{i+1} - x_{i-1})$ is the slope at grid point $x_i$, $\gamma = (x - x_i) / \Delta x$, and $h_{ab}$ are the Hermite cubic splines
\begin{subequations}
\begin{align}
h_{00}(\gamma) &= (1 + 2\gamma){(1 - \gamma)}^2 \\
h_{01}(\gamma) &= \gamma^2 (3 - 2\gamma) \\
h_{10}(\gamma) &= \gamma{(1 - \gamma)}^2 \\
h_{11}(\gamma) &= \gamma^2 (\gamma - 1).
\end{align}
\end{subequations}

\begin{figure}
	\includegraphics[width=\linewidth]{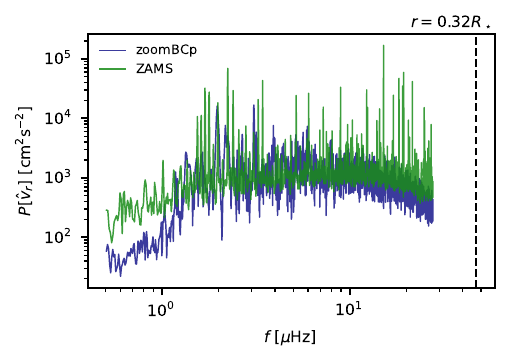}
	\caption{
		Power spectra computed from the \emph{ZAMS} full hemisphere simulation (\tabref{tab:simulations}), and the `wedge' simulation (\tabref{tab:wedge-simulations}) with periodic (\emph{zoomBCp}) boundary conditions in the angular direction.
		The vertical dashed line shows the Brunt-V\"ais\"al\"a frequency of the model, and the spectra are truncated at half the Nyquist frequency of the output dumps to eliminate spectral folding.
	}
	\label{fig:wedge-spectra}
\end{figure}

We performed simulations with online particles, and the parameters listed in \tabref{tab:wedge-simulations}.
These simulations are setup with a `wedge'-shaped domain, spanning from $\pi / 4$ to $3\pi / 4$ in the angular direction, and one pressure scale height above and below the convective boundary given in \tabref{tab:models}.
Both simulations in \tabref{tab:wedge-simulations} are the same except from the particles, so exhibit the same wave spectra.
We confirmed that the spectra generated in these simulations resembles that of the full hemisphere ZAMS simulation in \tabref{tab:simulations}.
The periodicity of these wedge simulations allows for \glspl*{igw} to continue `spiralling' (due to the dispersion relation, \eqref{eq:igw-dispersion}) from atop the convective core to the surface uninterrupted.
The differing domain and choice of angular boundary conditions could alter the behaviour of waves, so we compare the power spectra of a wedge and full hemisphere simulation in \figref{fig:wedge-spectra}.
The main differences between the spectra can be explained by the reduced size of the simulation domain.
High frequency g-modes in the full hemisphere simulation have radial wavelengths larger than the radial extent of the wedge domain, so these modes are forbidden in the zoomed simulation.
Likewise, the lowest frequency waves in the full hemisphere simulation are forbidden in the wedge domain due to its reduced angular extent limiting the horizontal wavelength of \glspl*{igw}.
This may contribute to the reduced power in low frequencies $f < f_{\rm conv}$ of the zoomed simulation.
Additionally, a slightly higher convective frequency associated with the smaller length scale of the convective region could play a part by inducing a shift of wave power towards higher frequencies.

\begin{table}
	\caption{
		Parameters for the additional simulations performed with online particle integration, and periodic boundary conditions.
	}
	\label{tab:wedge-simulations}
	\centering
	\begin{tabular}{llrrl}
		\hline \hline
		Model & $N_r \times N_\theta$ & $N_p$ $^a$ & $N_{\rm conv}$ & Interpolation \\
		\hline
		ZAMS & 276x336 & 243,605 & 120 & Linear \\
		ZAMS & 276x336 & 243,605 & 120 & Hermite (cubic) \\
		\hline
		\multicolumn{5}{l}{$^a$ Total number of particles in the simulation.} \\
	\end{tabular}
\end{table}

Particle trajectories have been computed for each combination of interpolation scheme (linear or Hermite) and offline/online time integration.
The particles are initialised only within the radiative zone with approximately 250,000 particles total, corresponding to three per MUSIC grid cell radially.
Particles are not initialised within $\pi/16$ of the angular boundaries, since this was needed for the simulations with reflective boundary conditions in \tabref{tab:simulations}.
The timestep $\Delta t$ for the online particles (\eqref{eq:timestep-heun}) is $120\si{\second}$ on average, versus a timestep of $9\times 10^3 \si{\second}$ for the post-processed particles.
For a reference wave of $10\si{\micro\hertz}$, this gives on average 833 timesteps per wave period for the online particle case.

\begin{figure}
	\includegraphics[width=0.5\linewidth]{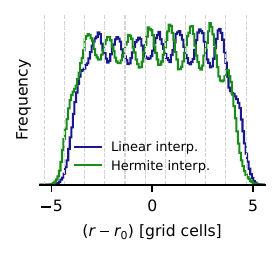}
	\includegraphics[width=0.5\linewidth]{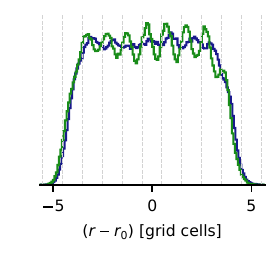}
	\caption{
		Histograms of particle positions at the end of the simulations listed in \tabref{tab:wedge-simulations}, computed over three wave periods.
		Only a small subset of the particles are visualised, restricted to within 4 radial MUSIC grid cells of $r=0.35R_\star$ so that the sub-cell `clumping' is visible.
		\emph{Left}: offline particles, computed in post-processing. \emph{Right}: online particles, computed in step with the hydrodynamical scheme.
	}
	\label{fig:wedge-histogram}
\end{figure}

\figref{fig:wedge-histogram} shows histograms of particle positions at the end of the simulations, using only a small radial subset of the particles so as to visualise the clumping artefact.
The histograms show similar distributions to that of the identically computed plot in the top-right panel of \figref{fig:particle-histogram} (computed from the full hemisphere ZAMS simulation), confirming the presence of the unwanted numerical artefact.
Since tracer particles should follow the mass, the accumulation of particles in the middle of cells is a spurious numerical artefact.
It is clear that both the interpolation and time integration have an impact on the artefact.
Linear interpolation leads to weaker artificial focusing of the particles than cubic spline interpolation, decreasing the variation in particle density by up to a factor of two.
The cubic splines possibly introduce local minima/maxima which should not be present in the velocity field, leading to increased clumping of the particles.
Integrating the particle trajectories online also leads to a weaker artefact by up to a factor of approximately two, and a decrease in computed diffusion coefficients by a similar factor.
Of course, the improvement from offline to online trajectory integration will depend on the time step $\Delta t$ in each.
However, neither change removed the clumping artefact, and running the simulations for longer would further amplify the effect.
Measuring small diffusion coefficients acting over evolutionary timescales requires robust particle tracking down to variations orders of magnitude smaller than those depicted in \figref{fig:wedge-histogram}, so factors of two in measurement sensitivity do not improve the results.
These tests show that neither online integration of particles nor a higher order interpolation scheme resolves the issues with particle integration.

